\documentclass{aa}  

\usepackage{graphicx}
\usepackage{txfonts}
\usepackage{multirow}
\usepackage{hyperref}
\usepackage{mathrsfs}
\usepackage{supertabular}
\usepackage{xcolor}

\begin{document}

   \title{The CARMENES search for exoplanets around M dwarfs}

   \subtitle{Chromospheric modeling of M2--3\,V stars with PHOENIX}

   \author{D.~Hintz\inst{1}
          \and  B.~Fuhrmeister\inst{1}
          \and  S.~Czesla\inst{1}
          \and  J.~H.~M.~M.~Schmitt\inst{1}
          \and  E.~N.~Johnson\inst{2}
          \and  A.~Schweitzer\inst{1}
          \and  J.~A.~Caballero\inst{3}
          \and  M.~Zechmeister\inst{2}
          \and  S.~V.~Jeffers\inst{2}
          \and  A.~Reiners\inst{2}
          \and  I.~Ribas\inst{4,5}
          \and  P.~J.~Amado\inst{6}
          \and  A.~Quirrenbach\inst{7}
          \and  G.~Anglada-Escud\'e\inst{6,8}
          \and  F.~F.~Bauer\inst{2}
          \and  V.~J.~S.~B\'ejar\inst{9,10}
          \and  M.~Cort\'es-Contreras\inst{3}
          \and  S.~Dreizler\inst{2}
          \and  D.~Galad\'{\i}-Enr\'{\i}quez\inst{11}
          \and  E.~W.~Guenther\inst{12,9}
          \and  P.~H.~Hauschildt\inst{1}
          \and  A.~Kaminski\inst{7}
          \and  M.~K\"urster\inst{13}
          \and  M.~Lafarga\inst{4,5}
          \and  M.~L\'opez~del~Fresno\inst{3}
          \and  D.~Montes\inst{14}
          \and  J.~C.~Morales\inst{4,5}
          \and  V.~M.~Passegger\inst{1}
          \and  W.~Seifert\inst{7}
          }

   \institute{Hamburger Sternwarte, University of Hamburg, 
              Gojenbergsweg 112, D-21029 Hamburg, Germany \\
              \email{dominik.hintz@hs.uni-hamburg.de}
         \and
              Institut für Astrophysik, 
              Friedrich-Hund-Platz 1, D-37077 Göttingen, Germany
         \and 
              Centro de Astrobiología (CSIC-INTA), ESAC, 
              Camino Bajo del Castillo s/n, E-28692 Villanueva de la Cañada, Madrid, Spain
         \and
              Institut de Ci\`encies de l'Espai (ICE, CSIC), Campus UAB, c/ de Can Magrans s/n, E-08193 Bellaterra, Barcelona, Spain
         \and
              Institut d'Estudis Espacials de Catalunya (IEEC), E-08034 Barcelona, Spain
         \and 
              Instituto de Astrof\'isica de Andaluc\'ia (CSIC), Glorieta de la Astronom\'ia s/n, E-18008 Granada, Spain 
         \and 
              Landessternwarte, Zentrum f\"ur Astronomie der Universit\"at Heidelberg, K\"onigstuhl 12, D-69117 Heidelberg, Germany 
         \and
              School of Physics and Astronomy, Queen Mary, University of London, 327 Mile End Road, London, E1 4NS, UK
         \and
              Instituto de Astrof\'{\i}sica de Canarias, c/ V\'{\i}a L\'actea s/n, E-38205 La Laguna, Tenerife, Spain
         \and
              Departamento de Astrof\'{\i}sica, Universidad de La Laguna, E-38206 Tenerife, Spain 
         \and
              Centro Astron\'omico Hispano-Alem\'an (MPG-CSIC), Observatorio Astron\'omico de Calar Alto, Sierra de los Filabres, E-04550 G\'ergal, Almer\'{\i}a, Spain 
         \and
              Th\"uringer Landessternwarte Tautenburg, Sternwarte 5, D-07778 Tautenburg, Germany 
         \and
              Max-Planck-Institut f\"ur Astronomie, K\"onigstuhl 17, D-69117 Heidelberg, Germany 
         \and
              Departamento de F\'{i}sica de la Tierra y Astrof\'{i}sica and UPARCOS-UCM (Unidad de F\'{i}sica de Part\'{i}culas y del Cosmos de la UCM), 
              Facultad de Ciencias F\'{i}sicas, Universidad Complutense de Madrid, E-28040, Madrid, Spain
        }

   \date{Received 6 December 2018; accepted dd Mm 2019}

  \abstract
  {
   Chromospheric modeling of observed differences in stellar activity lines is imperative to fully understand 
   the upper atmospheres of late-type stars.
   We present one-dimensional parametrized chromosphere models computed with the atmosphere code
   PHOENIX using an underlying photosphere of 3500 K. 
   The aim of this work is to model chromospheric lines of a sample of 50 M2--3 dwarfs 
   observed in the framework of the CARMENES, the Calar Alto high-Resolution search for M dwarfs with Exo-earths with 
   Near-infrared and optical Echelle Spectrographs, exoplanet survey. 
   The spectral comparison between observed data and models is performed in the chromospheric lines 
   of \ion{Na}{i}~D$_2$, H$\alpha$, and the bluest \ion{Ca}{ii}~infrared triplet line to obtain best-fit models
   for each star in the sample. We find that for inactive stars a single model with a VAL~C-like 
   temperature structure is sufficient to describe simultaneously
   all three lines adequately. 
   Active stars are rather modeled by a combination of an inactive and an active model, 
   also giving the filling factors of inactive and active regions. 
   Moreover, the fitting of linear combinations on variable stars yields relationships 
   between filling factors and activity states, indicating that
   more active phases are coupled to a larger portion of active regions 
   on the surface of the star.
   }


   \titlerunning{Chromospheric modeling of M2--3\,V stars with PHOENIX}
   \maketitle

\section{Introduction}
   Magnetic stellar activity comprises a zoo of phenomena affecting different layers of stellar atmospheres. 
   Magnetic activity is thought to be fundamental for the heating of the hot chromospheres and even hotter coronae of late-type stars, 
   which produce all of their high-energy ultraviolet and X-ray fluxes observed from these stars. 
   In addition, late-type stars are frequently planet hosts, and hence their activity is also recognized to have a crucial influence on the evolution of their planets, their
   atmospheric structure, and also possible life on their surfaces \citep[e.g.,][]{Segura2010AsBio..10..751S, France2016ApJ...820...89F, OMalley-James2017MNRAS.469L..26O}.
   
   In late-type stars, the chromosphere is the atmospheric layer between the photosphere and  transition region.  
   In the classical, one-dimensional picture, the atmospheric temperature reaches a minimum at the base of the chromosphere and then starts increasing outward. 
   Several heating processes, such as acoustic heating \citep{Wedemeyer2004A&A...414.1121W}, 
   back warming from the corona lying above the chromosphere and the transition region, and magnetic heating, are likely operating in the chromosphere. 
   The importance and magnitude of the individual proposed heating processes still remains 
   unsettled even in the case of the Sun.
   
   The chromosphere is the origin of a plethora of emission lines used to study its structure and physical conditions. 
   Solar images show the chromosphere to be highly inhomogeneous and constantly evolving \citep[e.g.,][]{Kuridze2015ApJ...802...26K}. 
   In the stellar context, there is the concept of the so-called basal chromospheric emission, since all stars show 
   some small extent of chromospheric activity which is referred to as basal chromospheric activity 
   \citep{Schrijver1987A&A...172..111S, Schrijver1989ApJ...341.1035S, Mittag2013A&A...549A.117M}.

   It is clear that a proper representation of such a chromosphere requires a dynamical three-dimensional approach. 
   For example, \citet{Uitenbroek2011ApJ...736...69U} give a thorough discussion about 
   the limitations of one-dimensional modeling even of stellar photospheres. 
   These authors concluded that the neglect of convective motions, nonlinearities of temperatures, and densities 
   in computing the molecular equilibrium and level populations as well as the 
   nonlinearities of the Planck function depending on the temperature may cause 
   inaccurate interpretations of the calculated spectra. On the other hand, 
   currently existing computer codes combining magnetohydrodynamic models 
   with realistic radiative transfer for the chromosphere (and also for the photosphere) still remain computationally 
   too costly to handle larger grids and cannot sensibly be juxtaposed to observations \citep{DeGennaroAquino2016phd}. Therefore, irrespective of its shortcomings, we consider the inferred chromospheric parameters from one-dimensional modeling
   useful, in particular, in a comparison among a sample of stars.
   
   Fundamental insights into the chromospheric structure can, however, already be obtained 
   based on static one-dimensional models with a parametrized temperature stratification.
   In the case of stars we are presumably looking simultaneously at the integrated emission 
   of many, spatially unresolved ``mini-chromospheres'', which may be described by a mean
   one-dimensional model; at least, such models can reproduce observed stellar spectra well for 
   M dwarfs \citep{Fuhrmeister2005A&A...439.1137F}. 
   On the other hand, the VAL~C model by \citet{Vernazza1981} is often considered the classical model for 
   the temperature structure in the photosphere, 
   the chromosphere, and the transition region of the average quiet Sun. 
   Up to now there are two approaches to model the chromospheres in late-type stars, viz., 
   scaling the VAL~C model \citep{Mauas1994A&A...281..129M, Fontenla2016ApJ...830..154F} 
   or parameterizing the chromosphere \citep{Short1997A&A...326..287S}. 
   Analyzing chromosphere models for a large stellar sample of homogeneous effective temperatures 
   gives the opportunity to detect whether and how the chromospheres are distinguished.

   In this study we use the state-of-the-art atmosphere code 
   PHOENIX\footnote{\url{https://www.hs.uni-hamburg.de/index.php?option=com_content&view=article&id=14&Itemid=294&lang=de}} \citep{Hauschildt1992JQSRT..47..433H, Hauschildt1993JQSRT..50..301H, Hauschildt1999JCoAM.109...41H} to compute 
   chromospheric model atmospheres based on a parametrized temperature stratification and obtain the resulting spectra. 
   These spectra are then compared to observed high-resolution spectra obtained with the CARMENES spectrograph. 
   Section~2 describes these observations and Sect.~3 highlights the model construction. 
   We compare the computed model spectra to a stellar sample of M2.0, M2.5, and M3.0 dwarfs observed by CARMENES 
   and search for the best-fit models to each star of the stellar sample in Sect.~4. 
   Moreover, we also fit linear model combinations to the spectra to improve 
   the modeling of the active stars where single models do not yield adequate fits. 
   By calculating model combinations we obtain filling factors illustrating the 
   coverage of inactive and active regions on the surfaces of the stars. 
   In Sect.~5 we present our results and conclusions.

\section{Observations}
   
   \subsection{CARMENES}
   The CARMENES spectrograph \citep[Calar Alto high-Resolution search for M dwarfs with Exo-earths with 
   Near-infrared and optical Echelle Spectrographs;][]{Quirrenbach2018SPIE10702E..0WQ} is a highly stabilized spectrograph 
   attached to the $3.5 \,$m telescope at the Calar Alto Observatory. 
   The visual channel (VIS) operates in the wavelength range 
   from $5200\,\AA$ to $9600\,\AA$, and the infrared channel (NIR) covers the range 
   between $9600\,\AA$ and $17\,100\,\AA$. 
   The spectral resolution of the VIS channel is about $R = 94\,600$ and 
   that of the NIR channel is $R = 80\,400$. 
   
   Since the start of the observations at the beginning of 2016, 
   CARMENES regularly observes a sample of $\sim$300 dM0 to dM9 stars in the context of
   its CARMENES survey \citep{Alonso-Floriano2015, Reiners2018A&A...612A..49R}. 
   The main goal of the survey is to find Earth-like planets in the habitable zone of 
   M dwarfs by measuring periodic signals in the radial velocities of the stars 
   with a precision on the order of $1 \, \mathrm{m} \, \mathrm{s}^{-1}$ 
   and long-term stability \citep{Reiners2018A&A...612A..49R, Ribas2018Natur.563..365R}. 
   The CARMENES survey sample is only magnitude limited in each spectral type. 
   Notably, no activity selection was applied. 
   
   The CARMENES spectra cover a wide range of chromospheric activity indicators 
   except for the classical \ion{Ca}{ii} H and K lines. 
   However, the \ion{Ca}{ii} infrared triplet (IRT) lines 
   can be observed, which were recently shown to be a good substitute of the blue \ion{Ca}{ii} H and K lines 
   \citep{Martinez-Arnaiz2011MNRAS.414.2629M, Martin2017A&A...605A.113M}. 
   Additional chromospherically active lines covered by the CARMENES spectrograph are H$\alpha$, 
   the \ion{Na}{i}~D lines, and the \ion{He}{i} D$_{3}$ line. 
   The shape of the different lines is well resolved, 
   including possible self-reversal in H$\alpha$ (in case of emission lines) 
   or the emission cores of \ion{Na}{i}~D. Traditionally, M dwarfs have been designated with 
   spectral class identifiers dM, dM(e), and dMe depending on whether H$\alpha$ 
   is in absorption, not detectable, or in emission in low-resolution
   spectra. We mainly use the indices of the \ion{Ca}{ii} IRT lines as activity
   indicators, since they simply fill in and go into emission without showing a complicated
   line profile like H$\alpha$. Nevertheless, we refer to the traditional designations in
   Table \ref{table_stars_basics} and examples of each class can be found in Fig. \ref{Star_sample_specs}.

   \subsection{Stellar sample}
   This study focuses on simulations and observations of the chromospheric properties of dM-type stars with an effective temperature
   of about $3500$~K, a surface gravity of $\log g = 5.0$, and solar metallicity $[\mathrm{Fe}/\mathrm{H}] = 0.0 \,$dex.
   These stars lie between the earliest M and the mid-M dwarfs. 
   While the stellar parameters can be fixed in the modeling, we have to allow some margin in the
   selection of targets from the CARMENES sample. 
   For this study, we selected all targets with stellar parameters fulfilling 
   $T_\mathrm{eff} = 3500 \pm 50 \, \mathrm{K}$, $\log g = 5.0 \pm0.2 \,$dex, 
   and $[\mathrm{Fe}/\mathrm{H}] = 0.0 \pm 0.3 \,$dex 
   as measured by \citet{Passegger2018A&A...615A...6P} using 
   PHOENIX photospheric models and also 
   given in the CARMENES input catalog Carmencita
   \citep{Carmencita_Caballero2016csss.confE.148C}. 
   Carmencita gathers information about the stars observed by CARMENES 
   from different sources, including the effective temperature, gravity, and metallicity. 
   We choose this range of effective temperature because it comprises a large number of M dwarfs, 
   some of which show H$\alpha$ in emission. 
   Below this effective temperature range the number of observed dM-type stars quickly decreases. 
   The stellar sample investigated in this work comprises 50 M~dwarfs 
   with spectral types between dM1.5 and dM3.5. 
   According to \mbox{\citet{Pecaut2013ApJS..208....9P}}\footnote{The effective temperatures for the spectral subtypes are given in 
   \url{http://www.pas.rochester.edu/~emamajek/EEM_dwarf_UBVIJHK_colors_Teff.txt}}, 
   dM2 and dM3 stars typically have effective temperatures at $3550$ and $3400\,$K. 
   Table~\ref{table_stars_basics} gathers basic data about our stellar sample including 
   Carmencita identifiers, names, spectral types, effective temperatures, surface gravities, and metallicities.

   \subsection{Data reduction} \label{data_reduction}
   The spectra for our sample stars provided in the CARMENES archive are
   reduced by the CARMENES data reduction pipeline 
   \citep{Caballero2016SPIE.9910E..0EC, Zechmeister2018A&A...609A..12Z}; note that 
   all wavelengths refer to vacuum. 
   Typical exposure times of the spectra were $800$, $900$, $1200$, and $1800\,$s, but there 
   are also several spectra with integration times lower than $500\,$s. 
   Since we want to compare line shapes to the model predictions 
   we require a minimum of $300\,$s of integration time 
   in order to exclude very noisy spectra. 
   
   To investigate the spectra, we corrected them for the barycentric velocity shift and also systematic 
   radial velocity shifts of the individual stars. 
   We did not apply a telluric correction since the chromospheric lines used in our study 
   are usually only weakly affected (H$\alpha$ and \ion{He}{i}~D$_{3}$). 
   However, spectra containing airglow signals in the cores of the \ion{Na}{i}~D lines are neglected 
   in the modeling. 
   Therefore we inspected by eye the spectra exhibiting possible 
   airglow contamination in the \ion{Na}{i}~D line cores (in the 
   wavelength range of $\lambda_\mathrm{NaD}\pm0.4\,\AA$) 
   and excluded those affected; 
   we only used spectra observed up until 21 December 2017. 
   The number of available and used spectra of our sample stars are listed in Table~\ref{table_stars_basics}; 
   most of the excluded spectra contain airglow signals.

   \begin{table*}[t]
   \caption{Basic information about the considered stars.$^a$}             
   \label{table_stars_basics}      
   \centering          
   \begin{tabular}{l l c c c c c c c c c}
   \hline\hline
   Karmn & Name & Sp. & Ref. & $T_\mathrm{eff}$ & $\log g$ & $[\mathrm{Fe}/\mathrm{H}]$ & $I_{\mathrm{H}\alpha}$ & $I_{\mathrm{Ca \, IRT}}$ & Available & Used  \\ 
    &  & type & (SpT) & $[\mathrm{K}]$ & $[\mathrm{dex}]$ & $[\mathrm{dex}]$ & $[\AA]$ & $[\AA]$ & spectra & spectra \\  
   \hline                    
  J00389+306 &          \object{Wolf 1056} &           dM2.5 &              AF15 &               3537 &                4.89 &                --0.04 &               1.74 &                0.63 &                12 &                  7                  \\ 
  J01013+613 &          \object{GJ 47} &               dM2.0 &              PMSU &                3537 &                4.92 &                --0.13 &               1.75 &                0.63 &                8 &                   1                  \\ 
  J01025+716 &          \object{BD+70 68} &            dM3.0 &              PMSU &                3478 &                4.92 &                0.00 &                1.73 &                0.62 &                107 &                 35                 \\ 
  J01433+043 &          \object{GJ 70} &               dM2.0 &              PMSU &                3534 &                4.91 &                --0.08 &               1.71 &                0.63 &                5 &                   4                  \\ 
  J02015+637 &          \object{G 244-047} &           dM3.0 &              PMSU &                3495 &                4.93 &                --0.05 &               1.73 &                0.62 &                17 &                  11                 \\ 
  J02442+255 &          \object{VX Ari} &              dM3.0 &              PMSU &                3459 &                4.96 &                --0.07 &               1.72 &                0.62 &                34 &                  19                 \\ 
  J03531+625 &          \object{Ross 567} &            dM3.0 &              Lep13 &               3484 &                4.94 &                --0.04 &               1.73 &                0.63 &                28 &                  28                 \\ 
  J06103+821 &          \object{GJ 226} &              dM2.0 &              PMSU &                3543 &                4.89 &                --0.05 &               1.76 &                0.63 &                19 &                  5                  \\ 
  J07044+682 &          \object{GJ 258} &              dM3.0 &              PMSU &                3469 &                4.94 &                --0.01 &               1.74 &                0.62 &                7 &                   7                  \\ 
  J07287-032 &          \object{GJ 1097} &             dM3.0 &              PMSU &                3458 &                4.95 &                --0.02 &               1.72 &                0.62 &                8 &                   8                  \\ 
  J07353+548 &          \object{GJ 3452} &             dM2.0 &              PMSU &                3526 &                4.93 &                --0.14 &               1.74 &                0.63 &                5 &                   2                  \\ 
  J09133+688 &          \object{G 234-057} &           dM2.5(e) &              Lep13 &               3545 &                4.93 &                --0.16 &               1.52 &                0.57 &                6 &                   2                  \\ 
  J09360-216 &          \object{GJ 357} &              dM2.5 &              PMSU &                3488 &                4.96 &                --0.14 &               1.73 &                0.63 &                2 &                   1                  \\ 
  J09425+700 &          \object{GJ 360} &              dM2.0(e) &              PMSU &                3511 &                4.91 &                --0.03 &               1.47 &                0.58 &                16 &                  12                 \\ 
  J10167-119 &          \object{GJ 386} &              dM3.0 &              PMSU &                3511 &                4.89 &                0.01 &                1.75 &                0.63 &                6 &                   3                  \\ 
  J10350-094 &          \object{LP 670-017} &          dM3.0 &              Sch05 &               3457 &                4.95 &                --0.03 &               1.74 &                0.62 &                5 &                   1                  \\ 
  J10396-069 &          \object{GJ 399} &              dM2.5 &              PMSU &                3524 &                4.91 &                --0.06 &               1.75 &                0.62 &                2 &                   2                  \\ 
  J11000+228 &          \object{Ross 104} &            dM2.5 &              PMSU &                3500 &                4.94 &                --0.10 &               1.74 &                0.63 &                47 &                  39                 \\ 
  J11201-104 &          \object{LP 733-099} &          dM2.0e &              Ria06 &               3540 &                4.97 &                --0.27 &               0.88 &                0.46 &                3 &                   3                  \\ 
  J11421+267 &          \object{Ross 905} &            dM2.5 &              AF15 &               3512 &                4.90 &                --0.02 &               1.75 &                0.63 &                113 &                 69                 \\ 
  J11467-140 &          \object{GJ 443} &              dM3.0 &              PMSU &                3523 &                4.87 &                0.06 &                1.76 &                0.62 &                5 &                   2                  \\ 
  J12230+640 &          \object{Ross 690} &            dM3.0 &              PMSU &                3528 &                4.87 &                0.03 &                1.77 &                0.63 &                80 &                  34                 \\ 
  J12248-182 &          \object{Ross 695} &            dM2.0 &              PMSU &                3476 &                4.98 &                --0.18 &               1.73 &                0.63 &                2 &                   2                  \\ 
  J14152+450 &          \object{Ross 992} &            dM3.0 &              PMSU &                3456 &                4.94 &                0.00 &                1.74 &                0.62 &                8 &                   6                  \\ 
  J14251+518 &          \object{$\theta$ Boo B} &           dM2.5 &              AF15 &               3512 &                4.92 &                --0.08 &               1.75 &                0.63 &                8 &                   5                  \\ 
  J15095+031 &          \object{Ross 1047} &           dM3.0 &              PMSU &                3480 &                4.93 &                --0.01 &               1.72 &                0.62 &                7 &                   7                  \\ 
  J15474-108 &          \object{LP 743-031} &          dM2.0 &              PMSU &                3515 &                4.96 &                --0.21 &               1.75 &                0.61 &                8 &                   6                  \\ 
  J16092+093 &          \object{G 137-084} &           dM3.0 &              Lep13 &               3455 &                4.98 &                --0.09 &               1.67 &                0.62 &                5 &                   5                  \\ 
  J16167+672N &         \object{EW Dra} &              dM3.0 &              PMSU &                3504 &                4.91 &                0.00 &                1.75 &                0.62 &                35 &                  23                 \\ 
  J16254+543 &          \object{GJ 625} &              dM1.5 &              AF15 &               3516 &                4.98 &                --0.27 &               1.75 &                0.63 &                32 &                  28                 \\ 
  J16327+126 &          \object{GJ 1203} &             dM3.0 &              PMSU &                3486 &                4.92 &                0.00 &                1.74 &                0.63 &                7 &                   5                  \\ 
  J16462+164 &          \object{LP 446-006} &          dM2.5 &              PMSU &                3505 &                4.92 &                --0.05 &               1.76 &                0.63 &                7 &                   6                  \\ 
  J17071+215 &          \object{Ross 863} &            dM3.0 &              PMSU &                3482 &                4.94 &                --0.05 &               1.73 &                0.62 &                7 &                   7                  \\ 
  J17166+080 &          \object{GJ 2128} &             dM2.0 &              PMSU &                3544 &                4.91 &                --0.10 &               1.75 &                0.63 &                8 &                   6                  \\ 
  J17198+417 &          \object{GJ 671} &              dM2.5 &              PMSU &                3499 &                4.93 &                --0.08 &               1.73 &                0.63 &                8 &                   8                  \\ 
  J17578+465 &          \object{G 204-039} &           dM2.5 &              AF15 &               3459 &                4.94 &                0.00 &                1.67 &                0.61 &                9 &                   8                  \\ 
  J18174+483 &          \object{TYC 3529-1437-1} &     dM2.0e &              Ria06 &               3515 &                4.96 &                --0.18 &               0.99 &                0.51 &                32 &                  32                 \\ 
  J18419+318 &          \object{Ross 145} &            dM3.0 &              PMSU &                3473 &                4.95 &                --0.06 &               1.71 &                0.62 &                7 &                   7                  \\ 
  J18480-145 &          \object{G 155-042} &           dM2.5 &              PMSU &                3500 &                4.94 &                --0.09 &               1.75 &                0.63 &                13 &                  5                  \\ 
  J19070+208 &          \object{Ross 730} &            dM2.0 &              AF15 &               3532 &                4.95 &                --0.21 &               1.69 &                0.63 &                10 &                  8                  \\ 
  J19072+208 &          \object{HD 349726} &           dM2.0 &              PMSU &                3535 &                4.94 &                --0.20 &               1.69 &                0.63 &                12 &                  9                  \\ 
  J20305+654 &          \object{GJ 793} &              dM2.5 &              PMSU &                3475 &                4.96 &                --0.08 &               1.64 &                0.62 &                25 &                  11                 \\ 
  J20567-104 &          \object{Wolf 896} &            dM2.5 &              PMSU &                3523 &                4.89 &                0.00 &                1.76 &                0.62 &                6 &                   6                  \\ 
  J21019-063 &          \object{Wolf 906} &            dM2.5 &              AF15 &               3521 &                4.90 &                --0.05 &               1.76 &                0.62 &                9 &                   5                  \\ 
  J21164+025 &          \object{LSPM J2116+0234} &     dM3.0 &              Lep13 &               3475 &                4.95 &                --0.05 &               1.75 &                0.62 &                21 &                  9                  \\ 
  J21348+515 &          \object{Wolf 926} &            dM3.0 &              PMSU &                3484 &                4.92 &                0.00 &                1.73 &                0.63 &                27 &                  20                 \\ 
  J22096-046 &          \object{BD-05 5715} &          dM3.5 &              PMSU &                3492 &                4.96 &                --0.01 &               1.76 &                0.62 &                41 &                  26                 \\ 
  J22125+085 &          \object{Wolf 1014} &           dM3.0 &              PMSU &                3500 &                4.92 &                --0.04 &               1.72 &                0.62 &                32 &                  29                 \\ 
  J23381-162 &          \object{G 273-093} &           dM2.0 &              PMSU &                3545 &                4.92 &                --0.13 &               1.78 &                0.63 &                8 &                   5                  \\ 
  J23585+076 &          \object{Wolf 1051} &           dM3.0 &              PMSU &                3470 &                4.94 &                0.00 &                1.73 &                0.62 &                16 &                  10                 \\ 
   \hline   
   \end{tabular}               
   \tablefoot{
   $^a$ Karmn is the Carmencita identifier.
   All effective temperatures ($\pm51\,$K), gravities ($\pm0.07\,$dex), and metallicities ($\pm0.16\,$dex) are measured by \citet{Passegger2018A&A...615A...6P}. 
   The indices $I_{\mathrm{H}\alpha}$ and $I_{\mathrm{Ca \, IRT}}$ are measured in this work.
   References: 
   AF15: \citet{Alonso-Floriano2015}; Lep13: \citet{Lepine2013}; 
   PMSU: \citet{Reid1995}; Ria06: \citet{Riaz2006AJ....132..866R}; 
   Sch05: \citet{Scholz2005A&A...442..211S}. 
   }
   \end{table*}

   \begin{figure}
   \centering
   \includegraphics[width=0.5\textwidth]{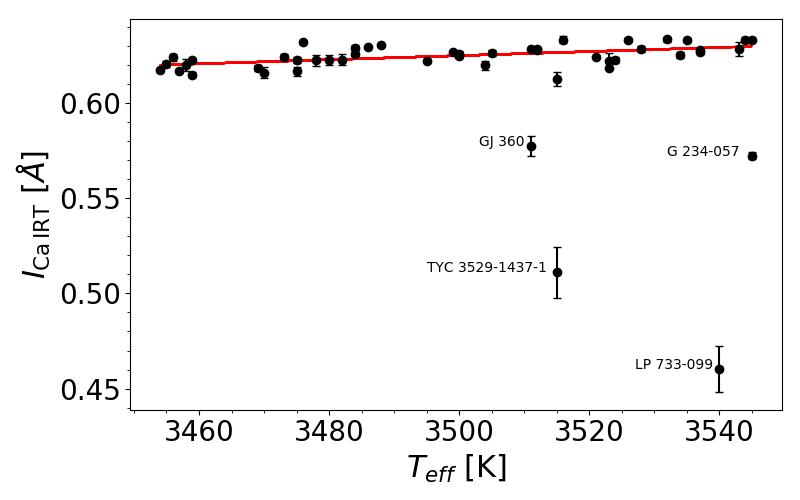}
      \caption{Average $I_\mathrm{line}$ in the \ion{Ca}{ii}~IRT line at $8500.35\,\AA$  
      against the $T_\mathrm{eff}$ of the investigated stars. 
      The error bars correspond to the standard deviations of the line indices of individual spectra of the respective stars.  
      The red line is a linear fit for the inactive stars 
      exceeding $I_\mathrm{Ca\, IRT} = 0.6\,\AA$. 
      The four active stars are highlighted by their names. 
              }
         \label{FigStellarSample}
   \end{figure}
   \begin{figure}
   \centering
   \includegraphics[width=0.5\textwidth]{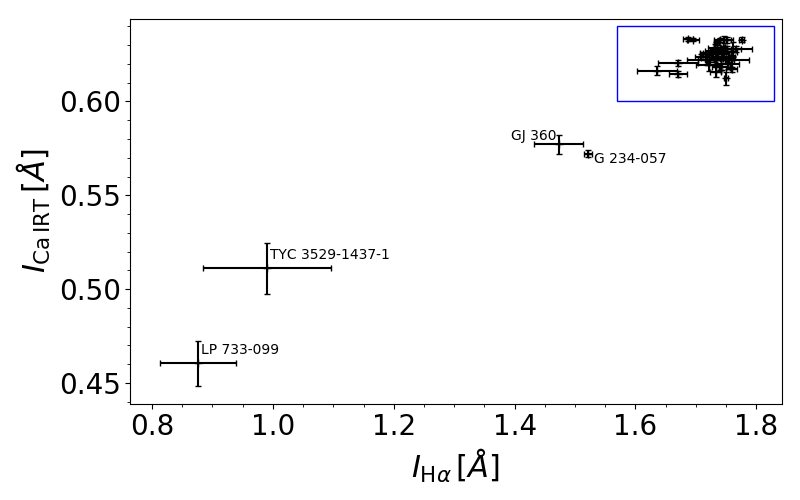}
   \includegraphics[width=0.5\textwidth]{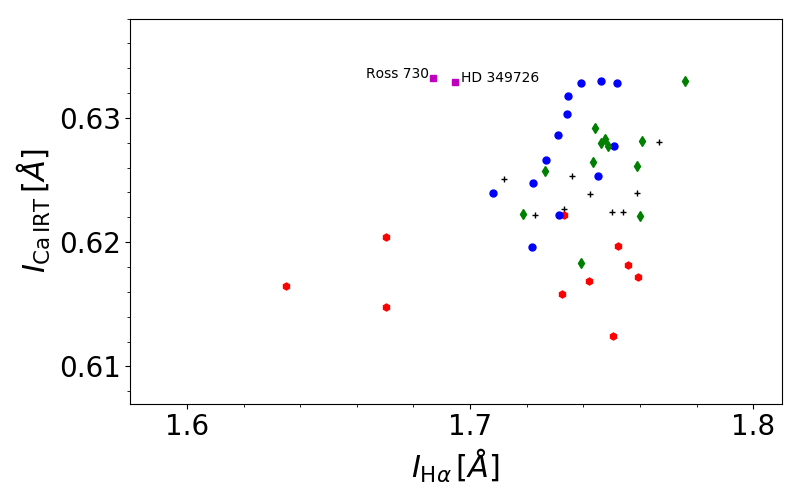}
      \caption{\textit{Upper panel:} Average $I_\mathrm{line}$ in the \ion{Ca}{ii}~IRT line at $8500.35\,\AA$ 
      against the average $I_\mathrm{line}$ in the H$\alpha$ line of the investigated stars. 
      The error bars correspond to the standard deviations of the line indices of individual spectra of the respective stars. 
      \textit{Lower panel:} Zoom of the range of the inactive stars without error bars (blue box in the upper panel). 
      The black pluses indicate the stars best fit by model \#079, red hexagons indicate best fit by model \#080, 
      blue circles by \#042, green diamonds by \#047, and magenta squares by \#029 (see Sect.~\ref{single}). 
      The model numbers and properties are given in Table~\ref{table_model_grid} in the appendix. 
      To improve clarity, the errors are faded out in the lower panel. 
              }
         \label{FigStellarSample_pewha_pewca}
   \end{figure}

   \subsection{Activity characterization of the sample stars} \label{act_char}
   
   The stars in the sample feature very different levels of stellar activity as can be easily seen 
   qualitatively in the H$\alpha$ line, which is an absorption line for most of our sample stars, but goes 
   into emission for four stars. 
   The state of activity can be characterized quantitatively by the line index of chromospheric emission lines. 
   Following the method of \citet{Fuhrmeister2018A&A...615A..14F} and \mbox{\citet{Robertson2016ApJ...832..112R}}, 
   the line index $I_\mathrm{line}$ is defined by 
   \begin{equation}
    I_\mathrm{line} = w \left( 1 - \frac{ \overline{F_\mathrm{line}} }
    { \overline{F_\mathrm{ref1}} + \overline{F_\mathrm{ref2}} } \right) 
   ,\end{equation}
   where $w$ is the bandwidth of the spectral line, and $\overline{F_\mathrm{line}}$, 
   $\overline{F_\mathrm{ref1}}$, and $\overline{F_\mathrm{ref2}}$ indicate the mean 
   flux densities of the spectral line and reference bands. 
   The line index corresponds to the pseudo-equivalent width. 
   
   In this paper, we investigated the \ion{Na}{i}~D, H$\alpha$, and \ion{Ca}{ii}~IRT lines. 
   These lines represent the most widely used chromospheric indicators in our wavelength range. 
   Another known chromospheric line covered by our wavelength range 
   is the \ion{He}{i}~D$_{3}$ line, which is seen in none of the spectra of our more inactive stars. 
   Of the \ion{Na}{i}~D doublet and the \ion{Ca}{ii}~IRT triplet lines, we only considered the bluest components because the remaining lines are influenced in the same manner by the chromospheric structure. 
   For the \ion{Ca}{ii}~IRT line we chose $w = 0.5\,\AA$ to be the width of the line band 
   centered at the (vacuum) wavelength of the line at $8500.35\,\AA$. 
   The reference band located at the blue side 
   of the Ca line is centered at $8481.33\,\AA$ with a half-width of $5\,\AA$, 
   and the central wavelength of the red band is $8553.35\,\AA$ with a half-width of $1\,\AA$ 
   to avoid telluric contamination for most radial and barycentric velocities. 
   The center of the H$\alpha$ line band is located at a wavelength of $6564.62\,\AA$ and we chose
   the width of the line band to be $1.6\,\AA$. 
   The reference bands are located at $6552.68\pm5.25\,\AA$ and $6582.13\pm4.25\,\AA$. 
   For the \ion{Na}{i}~D$_2$ line at $5891.58\pm0.2\,\AA$ the reference bands are 
   $5872.3\pm2.3\,\AA$ and $5912.0\pm2.0\,\AA$. 
   
   Fig.~\ref{FigStellarSample} shows an overview of the activity levels and the
   spread in effective temperature of the stellar sample. 
   Most of the stars are considered inactive indicated by the high \ion{Ca}{ii}~IRT index
   above $0.6\,\AA$, which should represent the average activity level of the star. 
   The more active stars in our sample are GJ~360 and \mbox{G~234-057} exhibiting line indices between $0.6\,\AA$ and $0.55\,\AA$ 
   (further on also called semi-active), 
   and even more active are \mbox{LP~733-099} and \mbox{TYC~3529-1437-1} with indices below $0.55\,\AA$. 
   There is a photospheric trend to higher indices for higher effective temperatures marked as the 
   linear fit in Fig.~\ref{FigStellarSample}. 
   This trend can also be found for the photospheric models by \citet{Husser2013} 
   varying the effective temperatures between $3400$ and $3600\,$K ($\log g = 5.0\,\mathrm{dex}$, 
   $[\mathrm{Fe}/ \mathrm{H}] = 0.0\,\mathrm{dex}$ and $[\alpha/\mathrm{Fe}] = 0.0\,\mathrm{dex}$). 
   We used the index of the \ion{Ca}{ii}~IRT line to specify 
   the different activity levels of the investigated stars, although 
   in many studies, such as in \citet{Fuhrmeister2018A&A...615A..14F}, the line index of H$\alpha$ is used to determine 
   the activity state. 
   While the H$\alpha$ line is a strong line very sensitive to activity changes,
   the drawback in using the H$\alpha$ line is its evolution with increasing stellar activity: this line first goes into absorption, then fills up the line core and eventually goes into emission 
   \citep[see, e.\,g.,][]{Cram_Mullan_1979ApJ...234..579C}. 
   Thus more active stars can exhibit the same H$\alpha$ line index as less active stars. 
   The \ion{Ca}{ii} IRT line only fills in and goes into emission while the
   activity level increases, making its index easier to interpret in terms of the activity state. 
   
\begin{figure}
   \centering
   \includegraphics[width=0.5\textwidth]{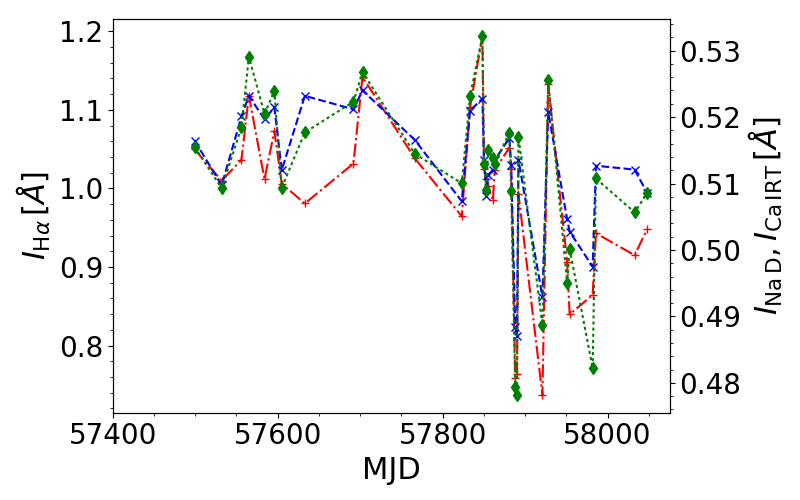}
      \caption{Time series of the indices $I_\mathrm{line}$ of the \ion{Na}{i}~D$_2$ (blue crosses), 
      H$\alpha$ (red pluses), and \ion{Ca}{ii}~IRT line (green diamonds) of \mbox{TYC~3529-1437-1}. 
              }
         \label{Fig_TYC_timeseries}
   \end{figure}

   In Fig.~\ref{FigStellarSample_pewha_pewca} we compare the average line indices of the \ion{Ca}{ii}~IRT line 
   and the H$\alpha$ line that are given in Table~\ref{table_stars_basics}. 
   The plot exhibits a good correlation between the 
   two indices with a Pearson correlation coefficient of $0.97$, while 
   the indices of the more inactive stars form an uncorrelated cloud. 
   At the highest \ion{Ca}{ii}~IRT line indices there 
   are two stars, Ross~730 and HD~349726, exhibiting lower H$\alpha$ indices;  we interpret these to be the most inactive
   stars of the sample being located in the branch where increasing activity means increasing
   H$\alpha$ line depth. 
   As error bars we plot the standard deviation of the line indices derived from individual spectra
   of the respective stars. 
   The scatter in the H$\alpha$ line index is obviously larger than that in the \ion{Ca}{ii}~IRT line. 
   So the H$\alpha$ line appears to be more sensitive for activity variations in the whole observation period. 
   This supports the choice for the \ion{Ca}{ii}~IRT line index as a robust estimate of the mean 
   activity level of a given star. 
   For further insight into the variability of individual stars we show the time series of the 
   three considered lines for \mbox{TYC~3529-1437-1} in Fig.~ \ref{Fig_TYC_timeseries} as
   one of the stars with the largest amount of variations. 
   Fig.~ \ref{Fig_TYC_timeseries} demonstrates the typical temporal sampling of the spectral time series, 
   which has been optimized for  planet searches. 
   The average sampling cadence is around 14 days. 
   Also times of different levels of activity can be identified: the star was more active at the end of the time series. 
   Since we do not want to average the variations of the lines or whole periods of enhanced activity, 
   we do not co-add the observations. Co-adding would certainly boost the 
   signal-to-noise ratio, but most stars also have signal-to-noise ratios above 50 in single
   spectra, which is sufficient for our analysis. 
   Therefore, we examine the spectrum with the median line index $I_\mathrm{Ca\, IRT}$ 
   as a representation of the median activity level of the star. 
   For the inactive stars the level of variation is much lower and an averaging would be possible
   in the sense of not mixing different activity states, but for consistency we treat these like the four active stars.

   \begin{figure*}
   \resizebox{\hsize}{!}
            {\includegraphics[width=0.5\textwidth]{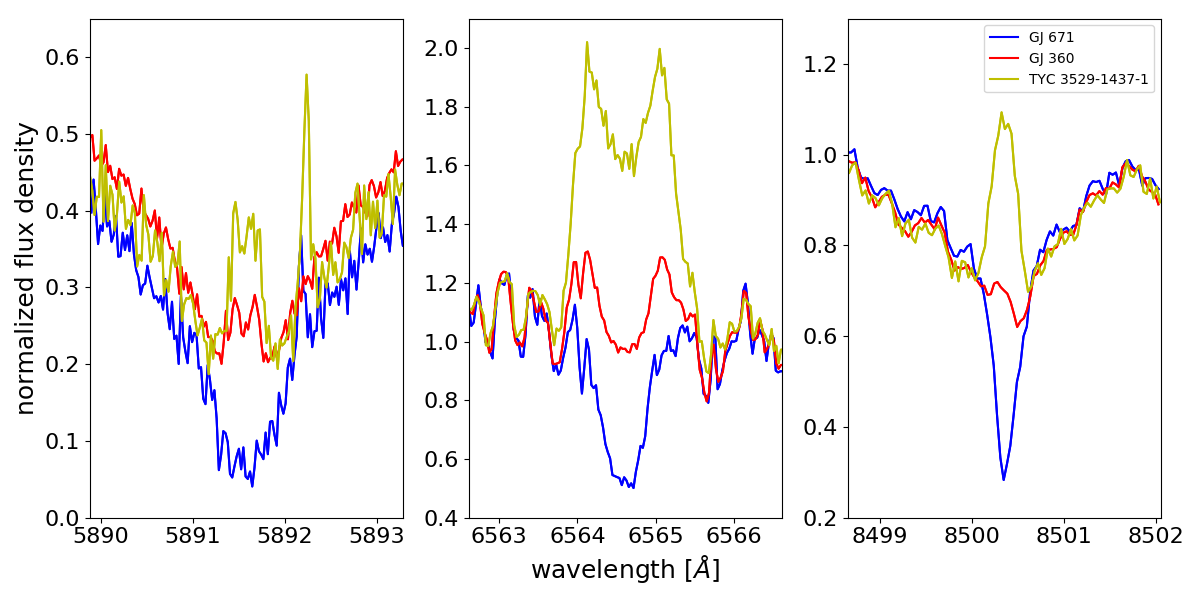}}
      \caption{Example spectra of one inactive (GJ~671, blue line), one semi-active 
               (GJ~360, red line), and one very active star 
               (\mbox{TYC~3529-1437-1}, yellow line) in the stellar sample 
               in the range between \ion{Na}{i}~D$_2$, H$\alpha,$ and the bluest \ion{Ca}{ii}~IRT line. 
               The peak right to the sodium line of \mbox{TYC~3529-1437-1} is an airglow line.
              }
         \label{Star_sample_specs}
   \end{figure*}
   
   In Fig.~\ref{Star_sample_specs} we show individual example spectra of one inactive (GJ~671) 
   and two active stars (GJ~360, \mbox{TYC~3529-1437-1}). 
   The inactive star exhibits absorption lines in 
   \ion{Na}{i}~D$_2$, H$\alpha$, and the bluest \ion{Ca}{ii}~IRT line. 
   For GJ~360 the spectral lines are in the transition from 
   absorption to emission, while the lines in the spectrum of \mbox{TYC~3529-1437-1} 
   are clearly in emission. 
   Thus we cover the bandwidth of rather inactive, semi-active, and very active M dwarfs in our sample.

\section{Model construction} \label{Models}
   The state-of-the-art PHOENIX code models atmospheres and spectra of a wide variety of 
   objects such as novae, supernovae, planets, and stars 
   \citep{Hauschildt1992JQSRT..47..433H, Hauschildt1993JQSRT..50..301H, Hauschildt1999JCoAM.109...41H}. 
   A number of PHOENIX model libraries covering M~dwarfs have been published. 
   For instance, \citet{Allard1995ApJ...445..433A} used local thermodynamic equilibrium (LTE) calculations 
   for dwarfs with effective temperatures of $1500\, \mathrm{K} \leq T_\mathrm{eff} \leq 4000 \, \mathrm{K}$ and 
   \citet{Hauschildt1999ApJ...512..377H} computed an even wider model grid in the temperature range between 
   $3000\,$K and $10000\,$K. 
   A recent library of PHOENIX model atmospheres was presented by \citet{Husser2013}.

   \subsection{Selection of a photospheric model}
   All of the above-mentioned libraries are exclusively concerned with the photospheres. 
   The first model calculations that investigated the chromospheres using PHOENIX were carried out by 
   \citet{Short1997A&A...326..287S} and \citet{Fuhrmeister2005A&A...439.1137F}. 
   These models extend the atmospheric range up to the transition region, but are still rooted in the
   atmospheric structure of the photosphere, which therefore, provides the basis for 
   our calculations. 
   
   In this paper, we adopt a photospheric model from the \citet{Husser2013} library as the underlying photosphere. 
   This model was calculated under the assumptions of
   spherical symmetry and LTE using 64 atmospheric layers. 
   The particular model atmosphere we adopted was computed for the parameters 
   $T_\mathrm{eff} = 3500\, \mathrm{K}$, $\log g = 5.0\,\mathrm{dex}$, 
   $[\mathrm{Fe}/ \mathrm{H}] = 0.0\,\mathrm{dex,}$ and $[\alpha/\mathrm{Fe}] = 0.0\,\mathrm{dex}$; 
   in Fig.~\ref{FigModelConstruction} (blue line) we show the temperature as a function of column 
   mass for this model photosphere. 
   Obviously, the temperature decreases continuously outward.

   \subsection{Chromospheric models} \label{chrom_models}
   We now extend the photospheric model following the approach of \citet{Short1997A&A...326..287S} and \citet{Fuhrmeister2005A&A...439.1137F}.
   In particular, we add three sections of rising temperature to the model photosphere, which represent the lower and upper chromosphere
   and the transition region. 
   To technically facilitate the extension, we increase the number of atmospheric layers in the model from originally 64 to 100 layers. 
   
\begin{figure}
   \centering
   \includegraphics[width=0.5\textwidth]{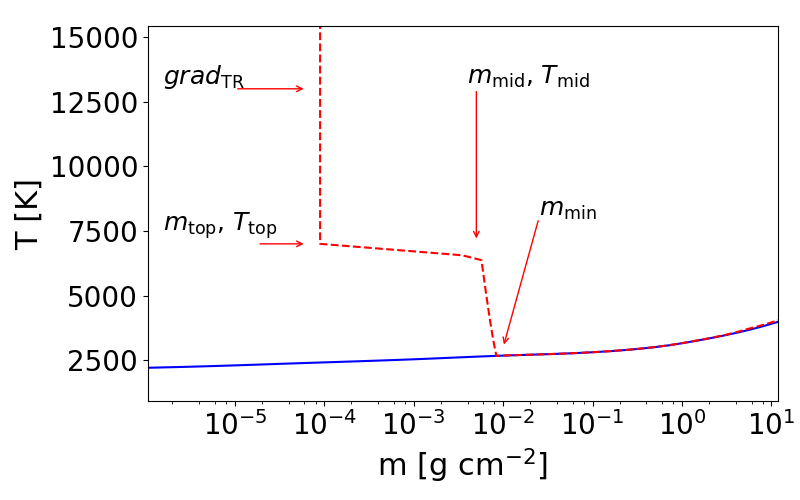}
      \caption{Photosphere model (blue solid line) at $T_\mathrm{eff} = 3500\,$K, $\log g = 5.0\,\mathrm{dex}$,
               $\left[\mathrm{Fe}/\mathrm{H}\right] = 0\,\mathrm{dex,}$ and 
               $\left[\alpha/\mathrm{Fe}\right] = 0\,\mathrm{dex}$ taken from the \citet{Husser2013} 
               library and an attached chromosphere model consisting of linear sections (red dashed line). 
               The arrows indicate the variable parameters of the chromosphere model. 
              }
         \label{FigModelConstruction}
   \end{figure}

   Our model chromosphere is described by a total of six free parameters. 
   The column mass density at the onset of the lower chromosphere, $m_\mathrm{min}$, 
   defines the location of the temperature minimum.
   The column mass densities and temperatures of the end points of the lower 
   ($\log m_\mathrm{mid}$, $T_\mathrm{mid}$) and the upper chromosphere ($\log m_\mathrm{max}$, $T_\mathrm{max}$),
   and the temperature gradient in the transition region, $grad_{\mathrm{TR}}$, given by 
   \begin{equation}
    grad_{\mathrm{TR}} = \frac{d\,T}{d\,\log\,m} = {\rm const.}
   \end{equation}
   as introduced by \citet{Fuhrmeister2005A&A...439.1137F}. 
   The maximum temperature of the transition region is fixed at $T_\mathrm{TR} = 98\,000\,$K. 
   The temperature rise segments in the lower, upper chromosphere, and transition region 
   are taken to be linear in the logarithm of the column mass density in our model. 
   This is a rough approximation to the structure of the VAL~C model. 
   The meaning of the individual parameters is also illustrated in 
   Fig.~\ref{FigModelConstruction}, where we show the temperature structure of the original photospheric model along with a 
   modified structure including the photosphere, chromosphere, and transition region. 
   
   We used this modified temperature structure as a new starting point for the PHOENIX calculations. 
   To account for the conditions in the chromosphere, 
   the atomic species of H~\textsc{i}, He~\textsc{i}~$-$~\textsc{ii}, C~\textsc{i}~$-$~\textsc{ii}, 
   N~\textsc{i}~$-$~\textsc{v}, O~\textsc{i}~$-$~\textsc{vi}, Na~\textsc{i}~$-$~\textsc{ii}, 
   Mg~\textsc{i}~$-$~\textsc{ii}, K~\textsc{i}~$-$~\textsc{ii}, and Ca~\textsc{i}~$-$~\textsc{iii} are computed 
   in non-LTE (NLTE) for all available levels taken from the database of 
   \citet{Kurucz1995all..book.....K}\footnote{\url{http://kurucz.harvard.edu/}}. 
   This applies to all species but He, for which we consulted CHIANTI~v4 \citep{Landi2006ApJS..162..261L}. 
   The PHOENIX code iteratively adapts 
   the electron pressure and mean molecular weight to reconcile these with the NLTE \ion{H}{i}/\ion{H}{ii} 
   ionization equilibrium. To that end, 
   it is necessary that the population of all the states of the NLTE species are 
   reiterated with the photoexciting and photoionizing radiation field, 
   which is particularly important 
   in the thin chromospheric and transition region layers. Moreover, the electron collisions 
   have to be taken into account. 
   A detailed comparison between LTE and NLTE atmospheric models is described by, for example,  \citet{Short2003ApJ...596..501S}. 
   Our calculations rely on the assumption of complete redistribution, 
   as PHOENIX does not yet support partial redistribution, which would provide
   a more appropriate treatment in large parts of the chromosphere. 
   However, its expected impact is largest for the Ly$\alpha$ line and resonance lines such as the 
   \ion{Ca}{ii} H and K line, which we also do not use in our study; 
   the \ion{Na}{i}~D lines are much less affected \citep{Mauas2000ApJ...539..858M}. 
   
   \begin{table}
   \caption{Parameter ranges for the chromosphere models.}             
   \label{parameter_ranges}      
   \centering
   \begin{tabular}{l c c}
   \hline\hline       
Parameter   & Minimum value   & Maximum value \\ 
 \hline 
$m_\mathrm{min}\,[\mathrm{dex}]$ & $-4.0$ & $-0.3$ \\ 
$m_\mathrm{mid}\,[\mathrm{dex}]$ & $-4.3$ & $-1.5$ \\ 
$T_\mathrm{mid}\,[\mathrm{K}]$   & $3500$ & $8000$ \\ 
$m_\mathrm{top}\,[\mathrm{dex}]$  & $-6.0$ & $-3.5$ \\ 
$T_\mathrm{top}\,[\mathrm{K}]$   & $4500$ & $8500$ \\ 
$grad_{\mathrm{TR}}\,[\mathrm{dex}]$     & $7.5$  & $10.0$ \\ 
   \hline   
   \end{tabular}
   \end{table}
   
   \begin{figure}[h]
   \centering
   \includegraphics[width=0.5\textwidth]{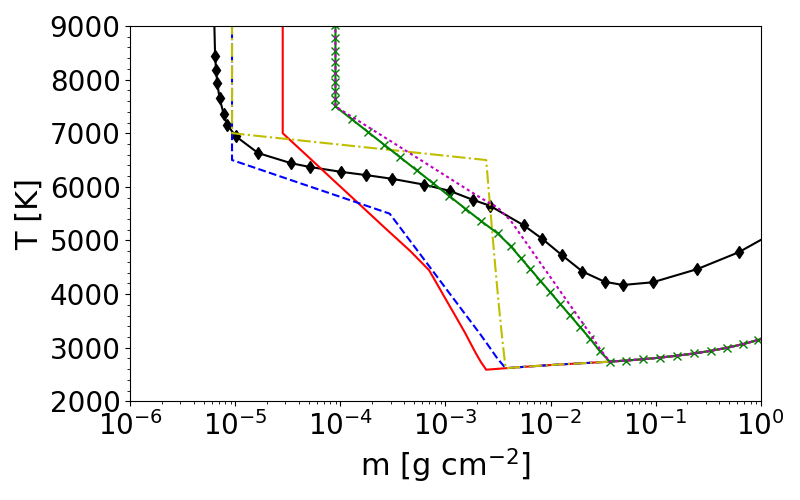} \\
   \includegraphics[width=0.5\textwidth]{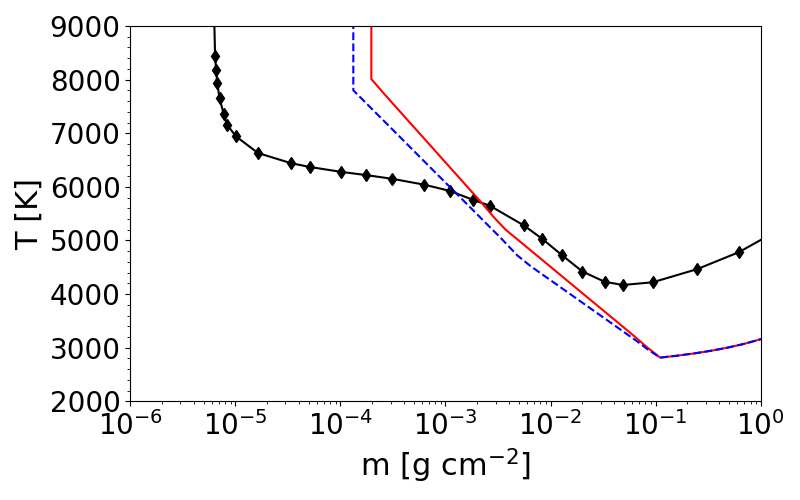}
      \caption{
               \textit{Upper panel:} Temperature structure of the best single-component fits 
               for the inactive stars as listed in Table~\ref{table_stars_models_best_mean} in the appendix. 
               Model \#029 is shown by the red solid line, \#042 by the blue dashed line, 
               \#047 by the yellow dash-dotted line, \#079 by the green crossed line, 
               and \#080 by the magenta dotted line. 
               \textit{Lower panel:} Same as in the upper panel for the active stars. 
               Model \#131 is plotted by the red solid line and \#136 by the blue dashed line. 
               The black solid diamond line shows the VAL~C model for the Sun \citep{Vernazza1981}. 
              }
         \label{model_grid_inactive_active}
   \end{figure}

   \subsection{Hidden parameters}\label{sec:hidden_pars}
   Besides the six parameters describing the temperature structure introduced in Sect.~\ref{chrom_models}, 
   there are further aspects of the model that can be modified and influence the solution, most importantly, 
   the microturbulence and the chosen set of NLTE lines. 
   
   While in our models the photospheric microturbulent velocity is set to $2\,\mathrm{km}\,\mathrm{s}^{-1}$, 
   in the chromosphere and the transition region the velocity 
   is set to half the speed of sound in each layer, but is not allowed to exceed 
   $20\,\mathrm{km}\,\mathrm{s}^{-1}$ which may otherwise happen in the transition region. 
   This follows the ansatz by \citet{Fuhrmeister2005A&A...439.1137F}. 
   We do not smooth the microturbulent velocity transition between photosphere and chromosphere but the conditions in the model lead to quasi-smooth transitions with increases between two layers   not exceeding $3\,\mathrm{km}\,\mathrm{s}^{-1}$.  
   Varying the microturbulent velocity leads to changes in the intensity as well as in the shape of the spectral lines 
   \citep{Jevremovic2000A&A...358..575J}. 
   The considered NLTE set is practically restricted by the computational effort. 
   While a more comprehensive NLTE set including, for example, higher O and Fe species would certainly be desirable to improve the synthetic spectra, 
   we consider the here-adopted NLTE set sufficient to model the chromosphere in the context of the investigated lines. 
   
   \begin{figure*}[t]
   \includegraphics[width=0.5\textwidth]{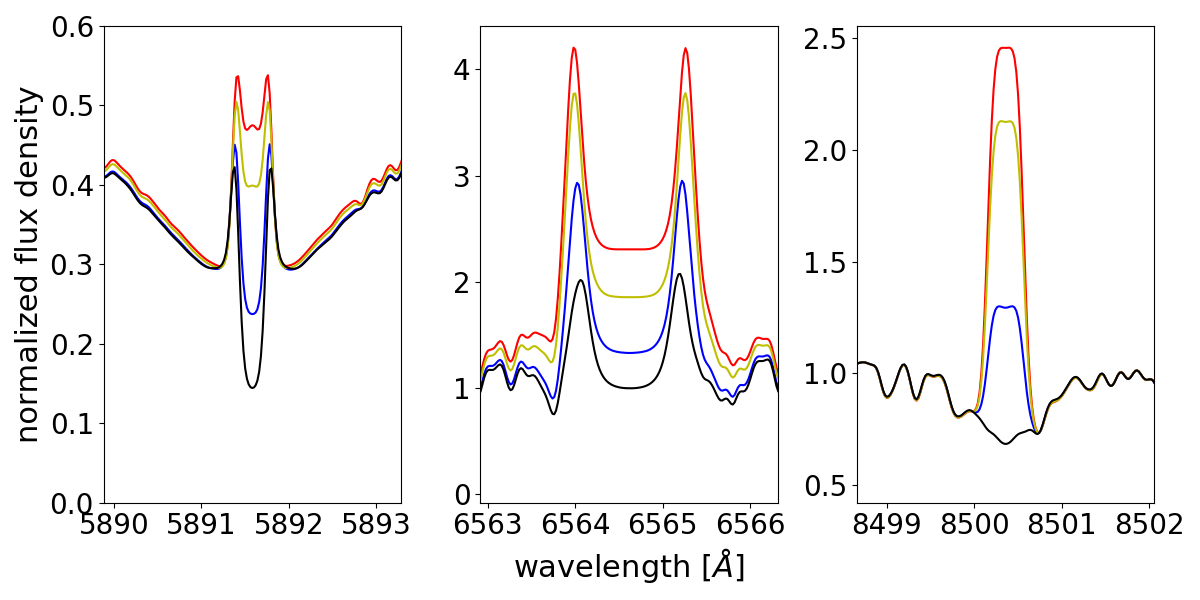}
   \includegraphics[width=0.5\textwidth]{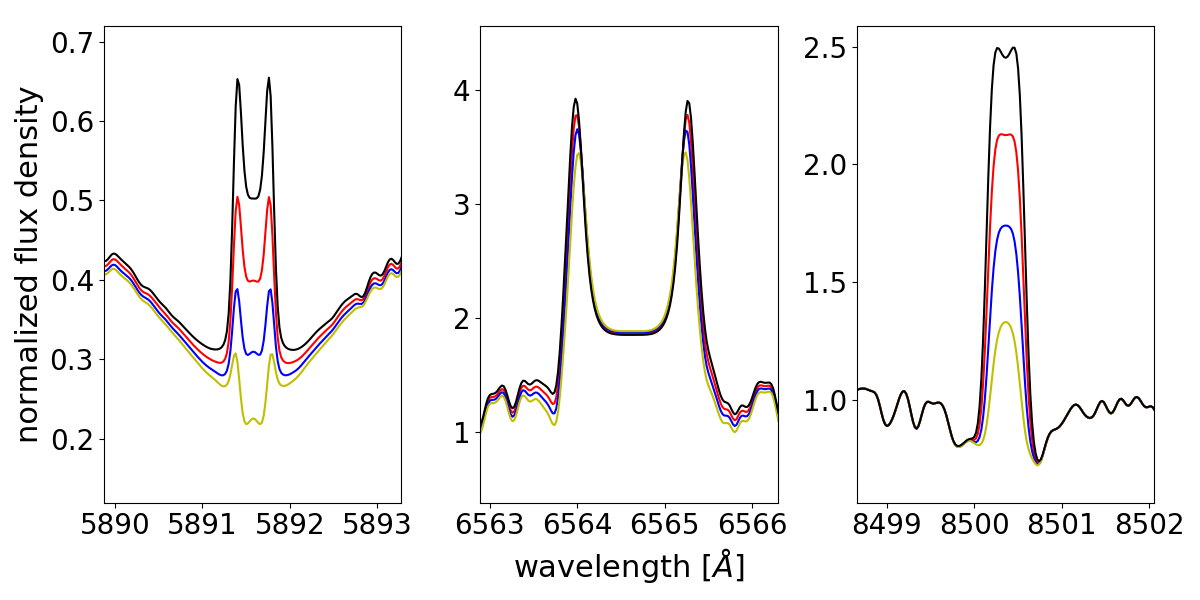} \\
   \includegraphics[width=0.5\textwidth]{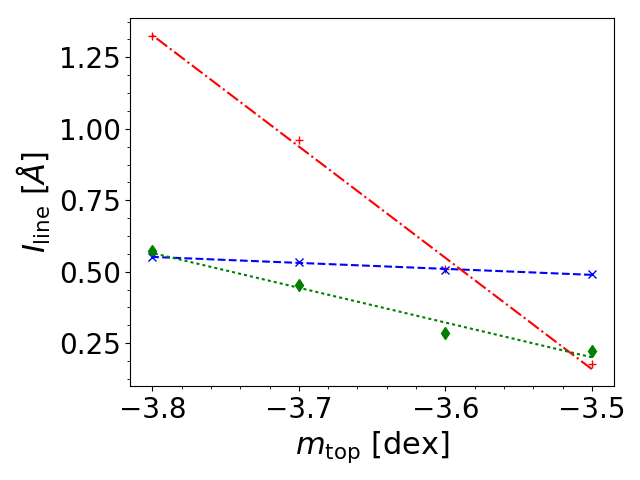}
   \includegraphics[width=0.5\textwidth]{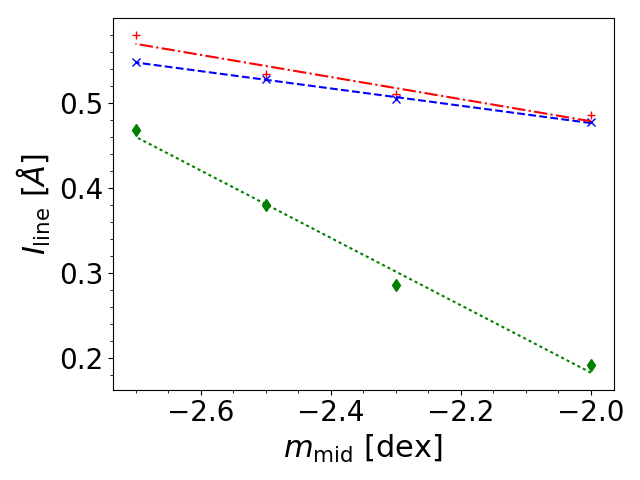}
      \caption{
      \textit{Top panel: 
      left-hand side:} Sequence of model spectra varying only the parameter $m_\mathrm{top}$ from $-4.0$ to $-3.5\,$dex, 
      the models are: 
      \#136 (black), \#137 (blue), \#138 (yellow), and \#139 (red). 
      \textit{Right-hand side:} Sequence for the parameter $m_\mathrm{mid}$ varying from $-2.7$ to $-2.0\,$dex, 
      models: \#121 (yellow), \#126 (blue), \#138 (red), and \#153 (black). 
      \textit{Bottom panel:} Corresponding values of $I_\mathrm{line}$ against $m_\mathrm{top}$ for \ion{Na}{i}~D (blue crosses), 
      H$\alpha$ (red pluses), and \ion{Ca}{ii}~IRT (green diamonds) for the top panels. 
      Additionally the linear fits are plotted: 
      the \ion{Na}{i}~D fit is plotted by the blue dashed line, the H$\alpha$ fit by the red dash-dotted line, and 
      the \ion{Ca}{ii}~IRT fit by the green dotted line. 
              }
         \label{series_lm_top}
   \end{figure*}

   \subsection{The model set} \label{results}
   Our chromospheric model is described by six free parameters, and 
   as a result of the rather large NLTE sets, each model calculation requires several dozens of CPU hours.
   While it may seem most straightforward to obtain a model grid homogeneously covering typical ranges for all free
   parameters, already a very moderate sampling of ten grid points per free parameter results in a
   grid with $10^6$ elements, 
   which results in too high a computational demand. 
   It is also clear that the large majority of the grid points are expected to result in spectra nowhere near the observations, which
   would be of little use in the subsequent analysis. 
   
   The challenge was therefore to identify reasonable parameter ranges and to explore these with
   a number of models. To that end, we first calculated limiting cases, such as models 
   simultaneously showing all spectral lines in absorption or emission, in other words inactive or active chromosphere models. 
   By visual comparison with the observed spectra, we subsequently identified the
   most promising parameter ranges to be explored further.
   
   A particularly interesting comparison is between models with a steep temperature rise in the lower 
   chromosphere and a plateau in the upper chromosphere, and models 
   with a shallow temperature increase in the lower chromosphere and a steeper increase in the 
   upper chromosphere. The first type of model is similar in structure to the VAL models used for the Sun. 
   
   The final set of models used in this study comprises 166 models with different model parameter configurations. 
   The parameters are varied in the ranges listed in Table~\ref{parameter_ranges}, and 
   the properties of the individual models are given in Table~\ref{table_model_grid} in the appendix. 
   The activity levels of the models are given by the line index $I_\mathrm{Ca\, IRT}$ 
   but, as shown in Sect.~\ref{sequences} and Sect.~\ref{sec:contri}, H$\alpha$ can behave differently compared to 
   the \ion{Ca}{ii}~IRT. 
   Figure~\ref{model_grid_inactive_active} shows an excerpt of the model set. 
   Three models have steep gradients in the lower chromosphere and shallow gradients in the upper chromosphere, and it turns out 
   these models represent more inactive states. 
   The other three models have the shallow gradients in the lower and the steep gradients in the upper chromosphere and are
   rather characteristic of active chromospheres.

   \subsection{Synthetic high-resolution spectra}   
   To compare our models to stellar spectra, we need densely sampled synthetic spectra.
   We calculated the synthetic spectra in the spectral ranges of $3900$--$4000\,\AA$, $4830$--$4890\,\AA$, 
   $5700$--$7000\,\AA$, and $8000$--$8800\,\AA$ 
   with a sampling of $\Delta \lambda = 0.005\,\AA$ per spectral bin. 
   These ranges comprise the chromospheric lines covered by CARMENES as discussed in Sect.~\ref{act_char} 
   and additionally the \ion{Ca}{ii} H and K, H$\epsilon$, and H$\beta$ lines to enable a comparison to spectra obtained by 
   other instruments such as the High Accuracy Radial velocity Planet Searcher \citep[HARPS;][]{Mayor2003Msngr.114...20M} in the context of further investigations. 
   For the comparison to the CARMENES spectra, we lowered the model spectral resolution to that of CARMENES 
   by folding with a Gaussian kernel of the approximated width. 
   The stellar rotational velocities are neglected because none of the sample stars is a fast rotator with 
   v $\sin i$ never exceeding $4\, \mathrm{km}\,\mathrm{s}^{-1}$ \citep{Reiners2017, Jeffers2018A&A...614A..76J}. 
   Even for the most active stars in our sample, 
   relatively long rotation periods are 
   not in contradiction with 
   their activity. \citet{Newton2017ApJ...834...85N} found the threshold between active and inactive stars
   (H$\alpha$ in emission or absorption) for early M dwarfs to be at a rotation period of about $30\,$days. 
   For two of our active dwarfs we have rotation period measurements by \citet{DiezAlonso2019A&A...621A.126D}: 
   GJ~360 has a period of $21\,$days and \mbox{TYC~3529-1437-1} has a period of $15.8\,$days. 
   These periods are consistent with the corresponding v $\sin i$ values and there is no need to assume that they are seen pole-on.

   \subsection{Sequences modeling the evolution of the chromospheric lines} \label{sequences}
   The interaction of the parameters determines whether the chromospheric lines 
   are in absorption or emission and determines the strengths and shapes of the lines. 
   In order to obtain an impression of how variations of a single parameter affect the line properties, we show in 
   Fig.~\ref{series_lm_top} the lines of \ion{Na}{i}~D$_2$, H$\alpha$, and the blue \ion{Ca}{ii}~IRT line of 
   different models only varying in the parameter $m_\mathrm{top}$ from $-4.0$ to $-3.5\,$dex (left-hand side) 
   and in $m_\mathrm{mid}$ from $-2.7$ to $-2.0\,$dex (right-hand side). 
   Changing a parameter of the upper chromosphere leads to stronger effects for H$\alpha$ compared to \ion{Na}{i}~D 
   and \ion{Ca}{ii}~IRT, although the \ion{Ca}{ii}~IRT is clearly more influenced than \ion{Na}{i}~D. 
   The line indices illustrated in the lower panel confirm this impression with 
   the gradients of the linear fits. 
   However, the line core of \ion{Na}{i}~D$_2$ is filled up while the peaks are increasing less. 
   The H$\alpha$ and \ion{Ca}{ii}~IRT lines completely go from absorption to emission, 
   and simultaneously the H$\alpha$ self-absorption becomes even stronger. 
   
   While H$\alpha$ is almost not influenced by the variation of $m_\mathrm{mid}$, the \ion{Ca}{ii}~IRT emission peak is 
   approximately four times higher above the continuum for $m_\mathrm{mid} = -2.0\,$dex than for $m_\mathrm{mid} = -2.7\,$dex. 
   The variation of \ion{Na}{i}~D is also visible in the spectra, but not as strong as for the \ion{Ca}{ii}~IRT line. 
   The strong self-absorption of \ion{Na}{i}~D leads to a smaller variation of the line index than for the \ion{Ca}{ii}~IRT. 
   This sequence suggests a weak relationship of the H$\alpha$ formation to the mid-chromosphere, 
   while the other two lines clearly depend on the structure of the transition between lower and upper chromosphere. 
   
   Furthermore, we also found the location $m_\mathrm{min}$ of the temperature minimum of the chromosphere 
   to be a decisive factor to determine 
   whether lines appear in absorption or emission. 
   More active lines are generated with a temperature minimum at higher density as found by \citet{Short1998A&A...336..613S} 
   who calculated a model grid for the chromospheric lines of H$\alpha$ and \ion{Na}{i}~D in five M dwarfs.

   \subsection{Flux contribution functions} \label{sec:contri}
\begin{figure}[t]
   \centering
   \includegraphics[width=0.5\textwidth]{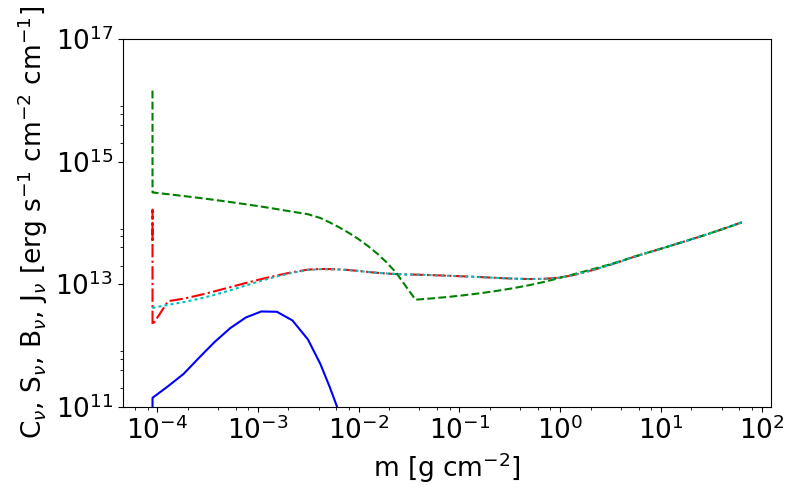} \\
   \includegraphics[width=0.5\textwidth]{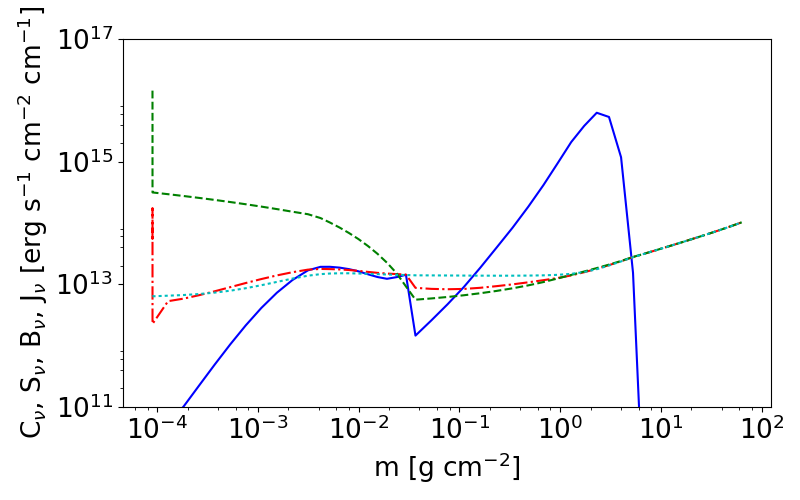} \\
   \includegraphics[width=0.5\textwidth]{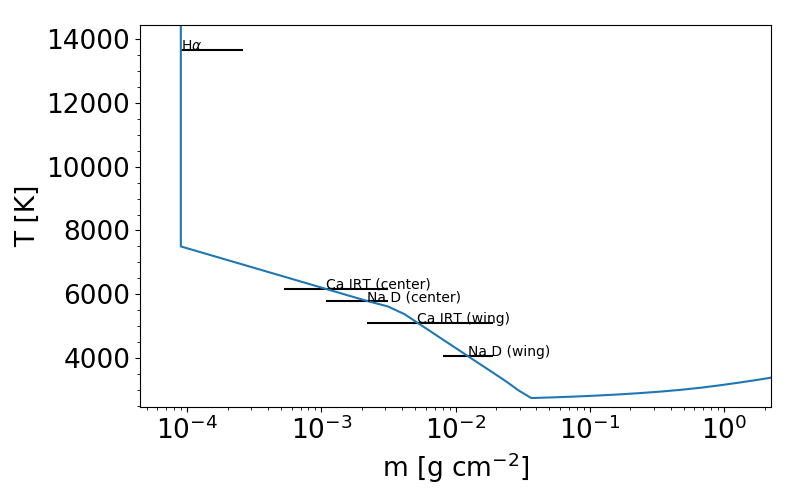}
      \caption{
      \textit{Upper and mid-panels:} Scaled flux contribution function $C_\nu$ (solid blue), 
      source function $S_\nu$ (dash-dotted red), Planck function $B_\nu$ (dashed green), and intensity $J_\nu$ (dotted cyan) 
      for $\lambda = 8500.36\,\AA$ representing the line core (upper panel) and $\lambda = 8500.54\,\AA$ in the line wing
      (mid-panel) of 
      the \ion{Ca}{ii}~IRT line in model \#080. The peak at
      high column mass in the contribution function of the line wing shows that the photospheric flux already dominates
      at this wavelength but there is still some chromospheric contribution. 
      \textit{Lower panel:} Temperature structure of model \#080 and the formation regions of \ion{Na}{i}~D$_2$, H$\alpha$, 
      and the bluest \ion{Ca}{ii}~IRT line. 
              }
         \label{Fig_contri}
   \end{figure}

   Performing sequences indicates that 
   modeling the chromospheric lines of \ion{Na}{i}~D$_2$, H$\alpha$, and the bluest \ion{Ca}{ii}~IRT line 
   turns out to be a challenge mainly due to different formation heights of the line wings and cores in the chromosphere. 
   Therefore, we investigated where the lines form in order to improve the understanding of the chromospheric structure and 
   to make restrictions for the calculations of the chromospheres. The
   PHOENIX code is capable of computing flux contribution functions following the
   concept of \citet{Magain1986A&A...163..135M} and \citet{Fuhrmeister2006A&A...452.1083F}. 
   The intensity contribution function $\mathscr{C}_I$ is defined by 
   \begin{equation}
    \mathscr{C}_I (\log m) = \mu^{-1} \, \ln 10 \, m \, \kappa \, S \, e^{-\tau/\mu},
   \end{equation}
   where $m$ is the column mass density, $\kappa$ the absorption coefficient, $S$ the source function, 
   $\tau$ the optical depth, and $\mu = \cos \theta$, with 
   $\theta$ as the angle between the considered direction and surface normal. 
   The flux contribution function $C$ is the intensity contribution function integrated over all $\mu$ 
   and it gives information on where the flux density of 
   every computed wavelength in the stellar atmosphere arises from. 
   Fig.~\ref{Fig_contri} illustrates the flux contribution function of the line core (upper panel) 
   and wing (mid-panel) of the bluest \ion{Ca}{ii}~IRT line for model \#080 (see Table~\ref{table_model_grid}). 
   Additionally, the source function $S_\nu$, Planck function $B_\nu$, and intensity $J_\nu$ are 
   given in the plots. 
   The core is formed nearly at the center of the upper chromosphere at $T = 6200 \,$K. 
   In the lower panel the temperature structure of the model is shown to visualize at which temperatures the lines are formed
   and the formation regions of \ion{Na}{i}~D$_2$, H$\alpha$ and the \ion{Ca}{ii}~IRT line are given. 
   The formation region is defined as the full width at half maximum around the maximum of the contribution function above the photosphere. 
   The \ion{Na}{i}~D$_2$ core is formed in the transition between the lower and upper chromosphere at $T = 5800 \,$K, 
   while the H$\alpha$ core is formed at the top of the chromosphere at $T = 13700 \,$K. 
   This is in agreement with the results from the best-fit model by \citet{Fuhrmeister2005A&A...439.1137F} 
   for AD~Leo in \ion{Na}{i}~D and H$\alpha$. 
   
\section{Results and discussion}
   
   \subsection{Comparing synthetic and observed spectra}\label{sec:comparison}
   To compare our synthetic spectra to observations we focus on 
   the \ion{Na}{i}~D$_2$, H$\alpha$, and the bluest \ion{Ca}{ii}~IRT lines. 
   Before carrying out the comparison, the flux densities of the observed and synthetic spectra were normalized to the mean value in the blue 
   reference bands of the respective chromospheric line (see Sect.~\ref{act_char}). 
   For stars with several available spectra, we used the spectrum with the median value of the $I_\mathrm{Ca\,IRT}$ line index for comparison. 

\begin{figure*}
   \resizebox{\hsize}{!}
            {\includegraphics[width=0.5\textwidth]{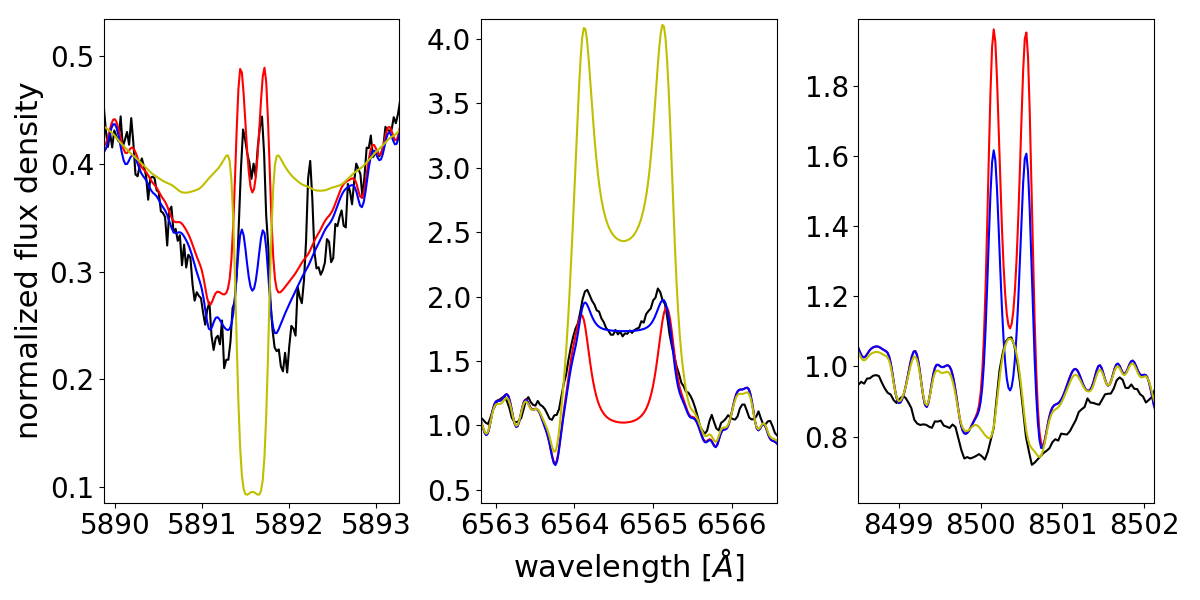}}
      \caption{
               Comparison of the observed median activity spectrum of \mbox{TYC~3529-1437-1} (black) as given by $I_\mathrm{Ca\, IRT}$ to the 
               model spectrum fitting best in a single line. Best-fit models using only \ion{Na}{i}~D$_2$ are denoted in red (model \#078), 
               using only H$\alpha$ in blue (model \#073), and  \ion{Ca}{ii}~IRT in yellow (model \#157). 
              }
         \label{inactive_obs_mods_3}
   \end{figure*}

   Our comparison is based on a least-squares-like minimization. 
   In particular, we consider the difference between the observed and
   synthetic spectra simultaneously in the wavelength ranges 
   $\lambda_\mathrm{Na D} \pm 0.2\,\AA$, $\lambda_\mathrm{H\alpha} \pm 0.8\,\AA$ and $\lambda_\mathrm{Ca IRT} \pm 0.25\,\AA$. These
   line bands cover the chromospheric cores of the respective spectral lines. 
   In Fig.~\ref{inactive_obs_mods_3} we show the best-fit models to the individual lines of \mbox{TYC~3529-1437-1} 
   irrespective of the other two lines.  
   In the case of the \ion{Ca}{ii}~IRT line, the best-fit model (yellow line) produces an almost perfect match to the observed line core, 
   however, a comparison of this model to the shape of \ion{Na}{i}~D and H$\alpha$ shows clear mismatches with regard to the data. 
   While the predicted H$\alpha$ line shows too strong an emission, 
   the core of the associated \ion{Na}{i}~D line profile is strongly absorbed, which is clearly at odds with the observation. 
   However, it is possible to obtain good fits for each of the three lines individually using appropriate models. 
   
   For physical reasons, the line band used for the H$\alpha$ line is nearly three times as wide as 
   that of the \ion{Ca}{ii}~IRT line and four times wider than that of the \ion{Na}{i}~D$_2$ line. 
   Moreover, the observed (and modeled) amplitude of variation in these line profiles is by far largest for the H$\alpha$ line. 
   Both factors boost the weight of the H$\alpha$ line profile in the minimization
   based on the summed $\chi^2$ values obtained in the three line bands as the objective. 
   This reflects a relative overabundance of data in the H$\alpha$ line and 
   makes it the dominant component in such a fit. 
   However, we  intend to find the model that provides the most appropriate representation
   of all three considered chromospheric lines simultaneously.  
  
   Therefore, we constructed a dedicated statistic, $\chi^2_m$, which we use as the objective function in our minimization.
   Specifically, we define
   \begin{equation}
       \chi^2_m = \chi^2_{\rm H\alpha} + w_{\rm Na}\, \chi^2_{\rm Na} + w_{\rm Ca}\, \chi^2_{\rm Ca} \; ,
   \end{equation}
   where $\chi^2_{\rm H\alpha}$, $\chi^2_{\rm Na}$, and $\chi^2_{\rm Ca}$ denote the sum of squared differences between model and observation obtained in the
   individual line bands. The weighting factors $w_{\rm Na}$ and $w_{\rm Ca}$ were adapted to give approximately the same
   weight to all three lines in the minimization. 
   To counterbalance the different widths of the line bands we increase the weighting of the \ion{Na}{i}~D$_2$
   and \ion{Ca}{ii}~IRT line bands by a factor of $3$. 
   To account for the different flux density scales of variation in the \ion{Na}{i}~D$_2$ 
   and \ion{Ca}{ii}~IRT lines, an additional factor of $4$ is added to further increase their weight. The factor of 4 was
   estimated based on the observed amplitude of variation of the H$\alpha$ line 
   compared to the variation of the \ion{Na}{i}~D$_2$ and \ion{Ca}{ii}~IRT lines. 
   We therefore end up with weighting factors of $w_{\rm Na} = w_{\rm Ca} = 12$. The resulting
   $\chi^2_m$ statistic establishes a relative order among the model fits, but cannot be used as a goodness-of-fit
   criterion.

   \subsection{Single-component fits} \label{single}
   In a first attempt, we directly compare our set of synthetic spectra with 
   the observations using the modified $\chi^2_m$ criterion defined above. 
   Since we normalize both the model and observation using the same reference band, no free parameters in 
   this comparison remain; however, we compare every observed spectrum to all spectra from our set of models.
   Table~\ref{table_stars_models_best_mean} in the appendix lists the results of the single model fits 
   for the whole stellar sample, i.\,e., the models with the lowest modified $\chi^2_m$ value. 
   An inspection of Table~\ref{table_stars_models_best_mean} shows that this approach yields reasonable fits with $\chi^2_m$ values
   between 1.8 and 4.04 for most stars, while four stars, the outliers in Fig.~\ref{FigStellarSample}, show only poor fits with
   $\chi^2_m$ values in excess of 10.
   
   \begin{figure*}[ht]
   \resizebox{\hsize}{!}
            {\includegraphics[width=0.5\textwidth]{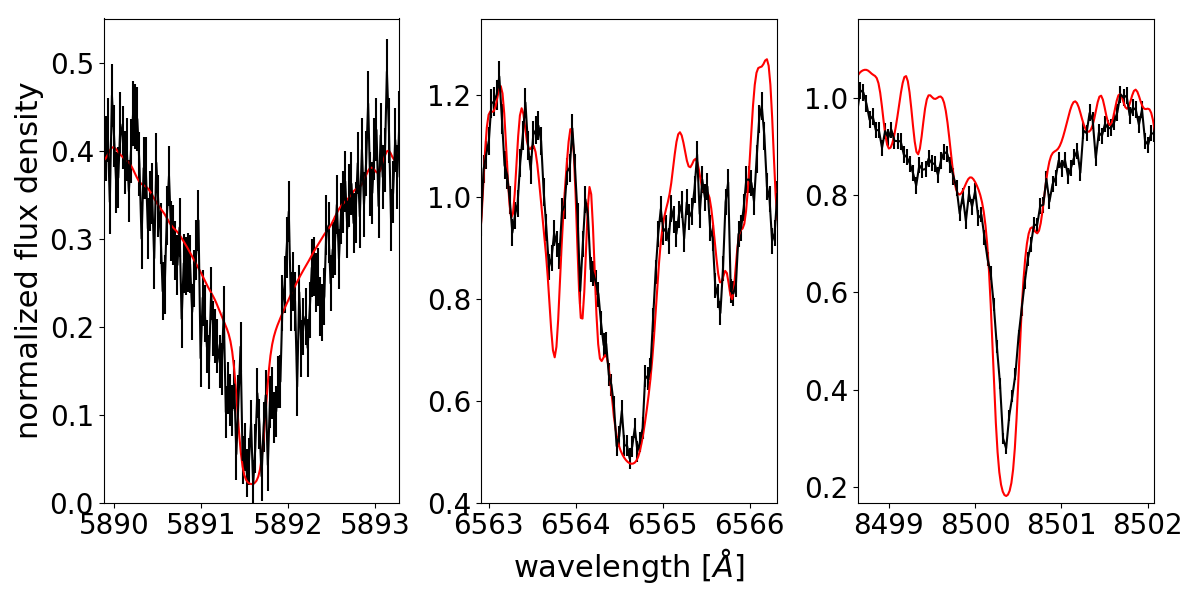}}
            \\
   \resizebox{\hsize}{!}
            {\includegraphics[width=0.5\textwidth]{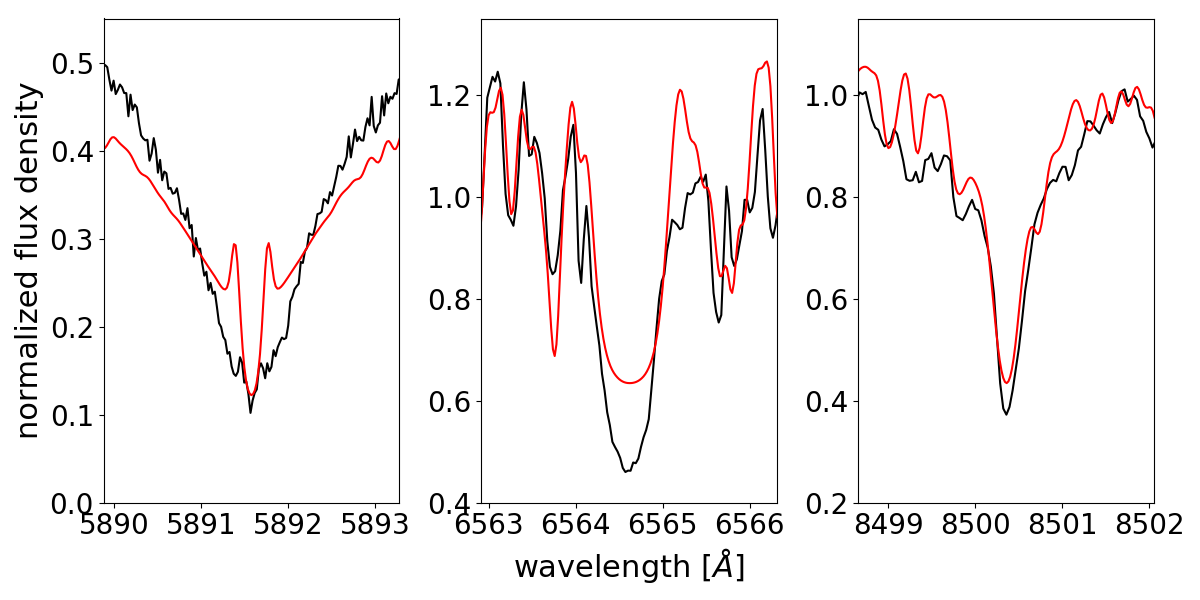}}
      \caption{
               Comparison of the best-fit model spectrum (red) to the observed spectrum (black) with median activity 
               as defined by $I_\mathrm{Ca\, IRT}$. 
               \textit{Upper panel:} GJ~671 and model \#042. \textit{Lower panel:} EW~Dra and model \#080. 
              }
         \label{inactive_obs_mods}
   \end{figure*}
   
   \subsubsection{Inactive stars}
   For the inactive stars in our sample interestingly a small subset of five models provides the best 
   fits, viz., the models \#029 (2 cases), \#042 (13 cases), \#047 (12 cases), \#079 (9 cases), and \#080 (10 cases). 
   In the lower panel of Fig.~\ref{FigStellarSample_pewha_pewca}, the distribution of inactive stars is shown in the plane 
   spanned by the H$\alpha$ and Ca~IRT activity indices. 
   The color coding identifies the best-fitting models from Table~\ref{table_stars_models_best_mean}. 
   While for most of the stars with $I_\mathrm{Ca\, IRT} < 0.625$ model \#080 is the best from our set,
   the spectra of the majority of stars exceeding this value are best represented by model \#042.
   The two stars Ross~730 and HD~349726, which we consider the most inactive in our sample 
   (see Sect.~\ref{act_char}), are best represented by model \#029. 
   The modified $\chi^2_m$ values range between $1.8$ and $4.5$ for the inactive stars. 
   
   As an example, we juxtapose the best-fitting models and observed spectra
   of the two inactive stars GJ~671 and EW~Dra in Fig.~\ref{inactive_obs_mods}. 
   In these models as well as in the observations all the considered lines appear in absorption. 
   We note that the observation of the sodium line of GJ~671 shows some telluric emission, however, sufficiently offset from the line core. 
   From our point of view, the \#042 provides a reasonable fit to all considered lines in the case of GJ~671.
   Despite some shortcomings, such as a too deep Ca~IRT line in the model and a somewhat too narrow sodium trough, all
   aspects of the data are appropriately reproduced by the model spectrum.  
   
   For the case of EW~Dra, the overall situation is less comfortable.
   The strongest deviation of model \#080 from the observed spectrum of EW~Dra is 
   the width of the central part of the \ion{Na}{i}~D$_2$ line. The model clearly produces a too wide line with
   a signature of a central fill-in and self-absorption clearly more pronounced than in the observation. We emphasize, however,
   that despite these differences, the observation shows a similar structure in the line core, yet of course less pronounced. 
   The line shapes of the H$\alpha$ and the \ion{Ca}{ii}~IRT line are well represented 
   by this model although in particular the observed H$\alpha$ line is deeper than that predicted by the model. 
   
   Although there are shortcomings, we conclude that the spectra of inactive stars can be appropriately represented by our models.
   Notably, all best-fit models for inactive stars are characterized by a temperature structure with a
   steeper temperature rise in the lower chromosphere and a more shallow rise in the upper chromosphere. 
   This is in fact not very different from the VAL~C model \citep{Vernazza1981} for the Sun, 
   which also shows a steeper temperature rise in the lower chromosphere and a plateau in the upper chromosphere. 
   
   \begin{figure*}[ht]
   \resizebox{\hsize}{!}
            {\includegraphics[width=0.5\textwidth]{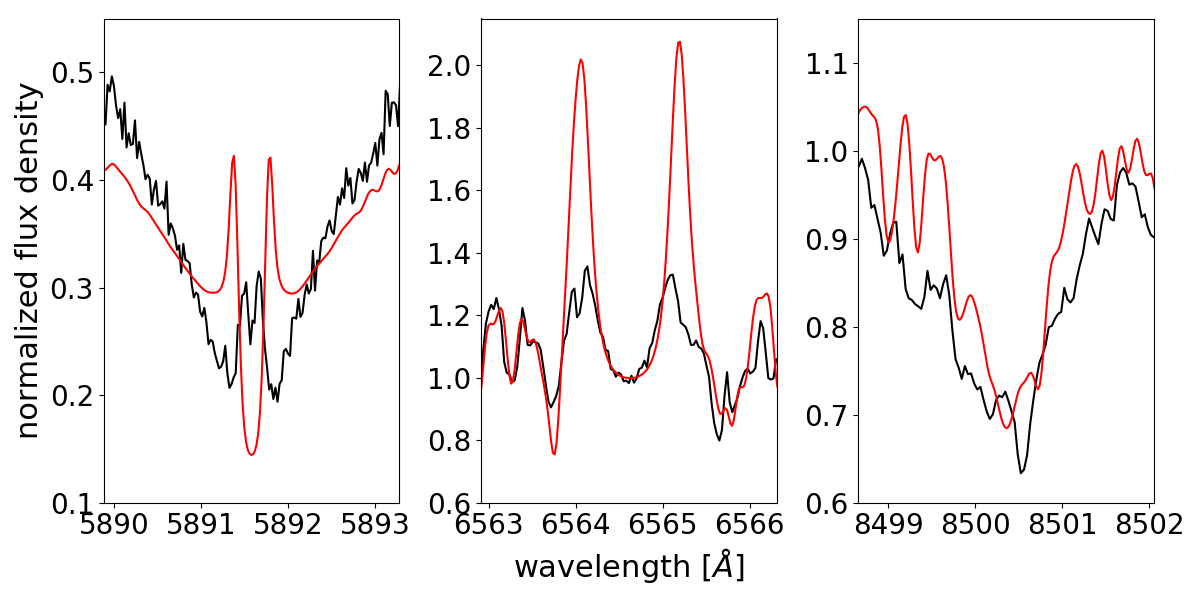}}
            \\
   \resizebox{\hsize}{!}
            {\includegraphics[width=0.5\textwidth]{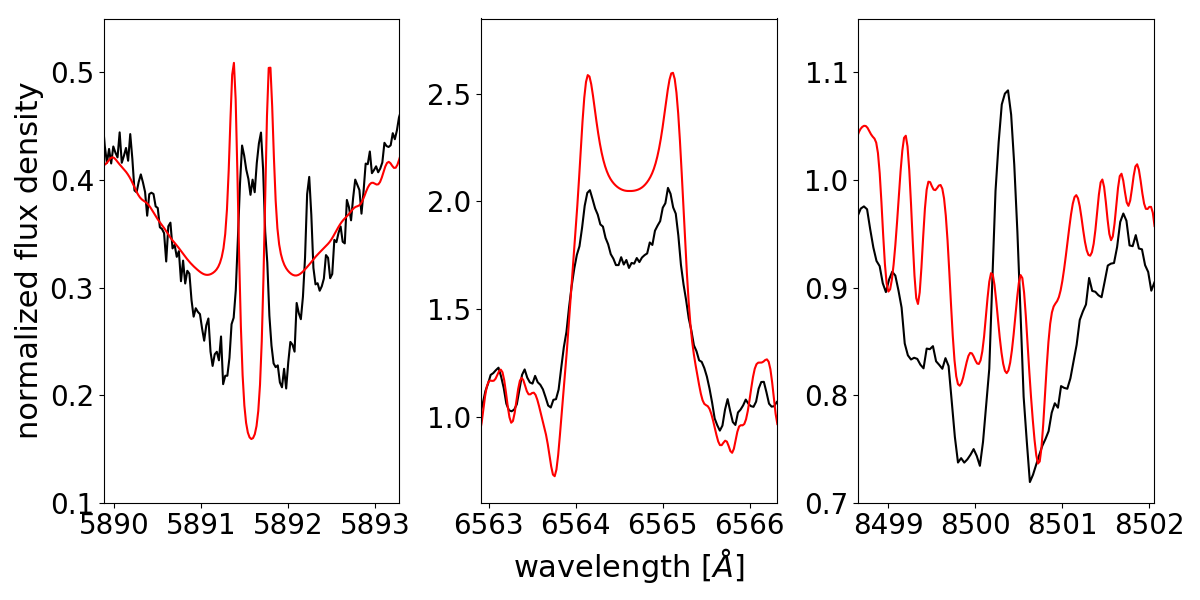}}
      \caption{
               Same as Fig.~\ref{inactive_obs_mods}. \textit{Upper panel:} GJ~360 and model \#136. \textit{Lower panel:} \mbox{TYC~3529-1437-1} and model \#131. 
              }
         \label{inactive_obs_mods_2}
   \end{figure*}

   \subsubsection{Active stars}
   For the active stars in our sample we only need two models that provide the best 
   fits, viz., the models \#131 (2 cases) and  \#136 (2 cases).  In
   Fig.~\ref{inactive_obs_mods_2} we show the observed spectra and best-fit models 
   for the more active stars GJ~360 and \mbox{TYC~3529-1437-1}, and clearly, 
   in both cases the modified $\chi^2_m$ values are more than twice as high as for
   any inactive star in the sample. 
   The observed \ion{Na}{i}~D line profiles show self-absorption in both stars, which is also predicted by the model, 
   but the width of the emission core is too broad for the models and the self-absorption is too strong. 
   The observed H$\alpha$ lines of both stars show clear self-absorption. 
   This aspect is reproduced by the models, although 
   for GJ~360 the model strongly overpredicts the self-absorption. 
   The overall H$\alpha$ line profile seems more
   appropriate for \mbox{TYC~3529-1437-1,} although the model yields a line that is too strong. 
   The shape of the \ion{Ca}{ii}~IRT line in the model is roughly reproduced in GJ~360; 
   the model for \mbox{TYC~3529-1437-1} shows considerable self-absorption, which is not observed. 
   In particular for the more active stars, it is
   hard to fit all three lines simultaneously by one model. 
   
   The column mass densities of the temperature minima $m_{\rm min}$ of the best-fit models of the active stars 
   are obviously higher compared to the models of the inactive stars. 
   Both models have a density of at least $\Delta m_{\rm min} = 0.5\,$dex higher than the found inactive models. 
   In notable contrast to our results for the inactive stars, the best-fit models for the spectra of the more active
   stars tend to show a more shallow rise in the lower chromosphere and a steeper rise in the upper chromosphere. 
   It is possible to increase the flux of the H$\alpha$ and \ion{Na}{i}~D$_2$ lines by only moving the chromosphere 
   to increasing densities, but the shape 
   of the chromosphere needs to be changed to reconcile a more active chromosphere with 
   the \ion{Ca}{ii}~IRT line. 
   Although this result has to be viewed against the background of the shortcomings of the fit, we tentatively identify this reversal of gradients
   as a characteristic change in the chromospheric temperature structure associated with a change from an inactive to an active chromosphere. 
   
   \begin{figure*}
   \resizebox{\hsize}{!}
            {\includegraphics[width=0.5\textwidth]{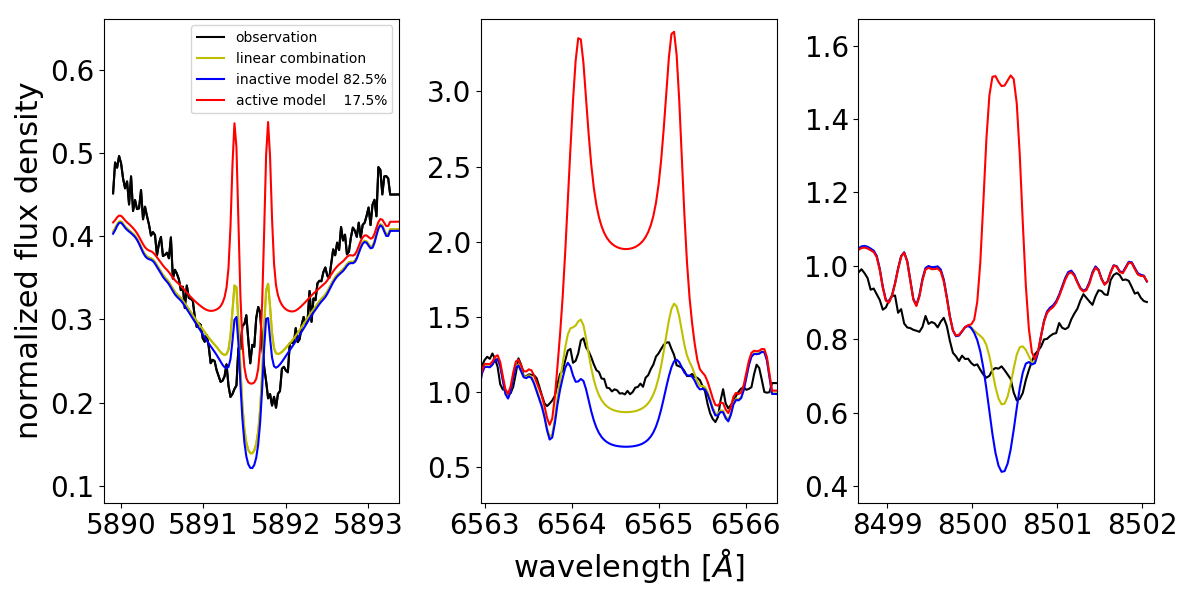}}
            \\
   \resizebox{\hsize}{!}
            {\includegraphics[width=0.5\textwidth]{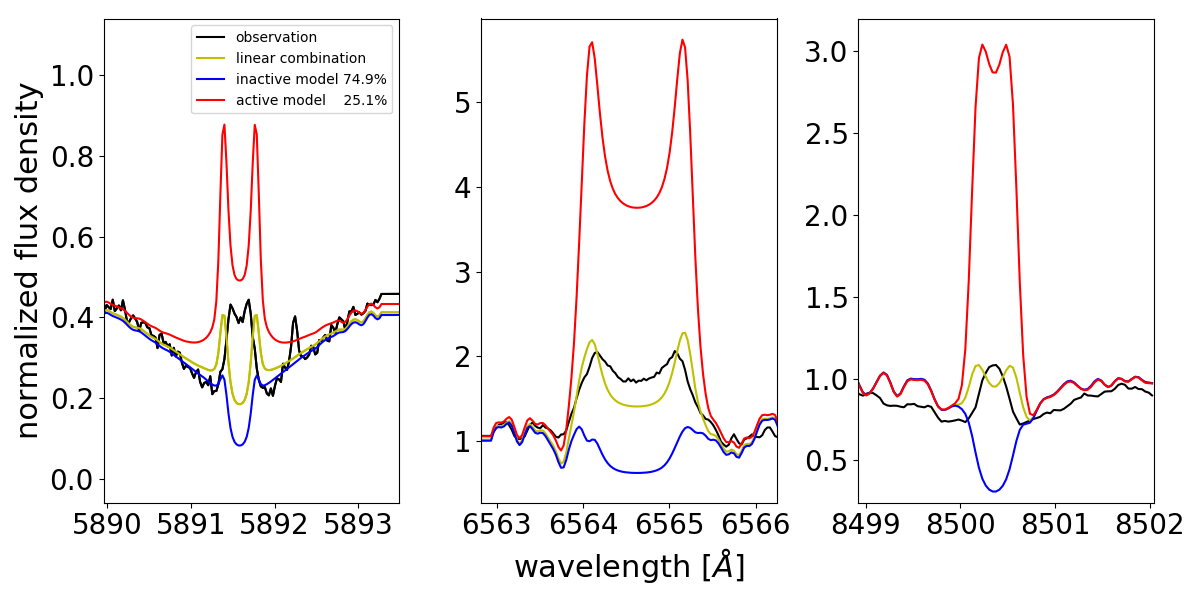}}
      \caption{
      Comparison of the observed median activity spectrum to 
      the best linear-combination fits for GJ~360 (\textit{upper panels}, inactive model \#080 and active model \#132) and 
      \mbox{TYC~3529-1437-1} (\textit{lower panels}, inactive model \#079 and active model \#149). 
      The line of \ion{Na}{i}~$\mathrm{D}_2$ and the bluest \ion{Ca}{ii}~IRT line 
      are weighted by a factor of $12$, respectively, compared to H$\alpha$ as described in Sect.~\ref{sec:comparison}. 
              }
         \label{model_lc2_active}
   \end{figure*}

   \subsection{Linear-combination fits} \label{lin_combs}
   The solar chromospheric spectrum is well known for varying across the solar disk, in particular,
   when active regions and quiet regions are compared. 
   To account for this fact, some studies using one-dimensional static models have combined models applying a filling factor. 
   This method was first applied using models with a thermal bifurcation 
   by \citet{Ayres1981ApJ...244.1064A} to account for observations of molecular CO in the solar spectrum. 
   However, it cannot be explained by chromospheric models alone, which have a temperature minimum well above 
   a temperature allowing for CO. Such studies normally use linear combinations 
   of a photospheric (or inactive chromospheric) and an active chromospheric model 
   and have also been used for flare modeling \citep{Fuhrmeister2010A&A...511A..83F}. 
   Classically this treatment is also known as 1.5D modeling, since it tries 
   to account for the inhomogeneity of the stellar atmospheres \citep{Ayres2006ApJS..165..618A}. It relies on the assumption 
   that the individual components are independent 
   and influence each other neither radiatively nor collisionally via NLTE. This assumption 
   would be satisfied if these regions were optically thick, which they are most probably
   not, but neither are they too optically thin. Therefore, we consider the assumption to be
   approximately satisfied. 
   
   We also adopt this approach in an attempt to
   improve especially the spectral fits of the more active dwarfs in our sample.
   In particular, we consider a linear combination of an inactive and an active spectral model, and determine 
   a best-fit filling factor by again minimizing the modified $\chi^2_m$ value. 
   This analysis is done for the whole stellar sample from Table~\ref{table_stars_basics}, and
   Table~\ref{table_stars_models_best_combined_8535} summarizes the results. 
   The latter table contains the filling factors of the inactive and active component 
   and the associated modified $\chi^2_m$ values. 
   Although these values do not represent classical $\chi^2$ value, they
   can be compared with those of the single model fits (see Table~\ref{table_stars_models_best_mean}). 
   As expected, the modified $\chi^2_m$ values improve for every star when a second component is added to the model. 
   Table~\ref{table_stars_models_best_combined_8535} also lists the differences of the 
   \ion{Ca}{ii}~IRT line indices ($\Delta I_\mathrm{Ca\, IRT}$) 
   between the inactive and active model in the best combination fits 
   giving an impression of the contrast in the respective combinations. 
   
   \subsubsection{Inactive stars}
   In the spectral fits of the inactive sample stars, we start with the best-fit single-component model
   as the first component. We then add every model of our model set, determine a best-fit filling factor, 
   and finally, choose the model combination providing the minimum modified $\chi^2_m$. 
   The thus determined second model component mostly also exhibits the \ion{Na}{i}~D$_2$, H$\alpha$ 
   and the \ion{Ca}{ii}~IRT lines in absorption. 
   Although the combination leads to better matching of the line shapes, 
   the overall improvement obtained for the inactive stars remains moderate.

   \subsubsection{Active stars}
   The spectra of the active stars could not be reproduced well with our single-component fits.
   Therefore, we do not consider the best-fit single model component as a good starting point for the fit
   as for the inactive stars and, instead, apply the following procedure.    
   As the inactive spectral model component, we subsequently tested the five models, which produced the best fits
   for the inactive stars in the single-component approach (see Sect.~\ref{single}). These were then tentatively combined
   with all other models from our set and the filling factor determined. Again, we finally opted for that combination
   providing the minimum modified $\chi^2_m$ value.  
   
   The best $\chi_m^2$ value is obtained by the combination with models 
   with differences regarding the temperature structure in comparison to the found inactive models. 
   The temperature gradient 
   of the lower chromosphere is shallower and that of the upper chromosphere is steeper, 
   which is the reverse case for the five best models of the inactive stars. 
   While the active models of \mbox{G~234-057} and GJ~360 
   have higher gradients in the upper than in the lower chromosphere, in the active model 
   of \mbox{LP~733-099} and \mbox{TYC~3529-1437-1} the gradients 
   of the lower and upper chromosphere are equal, meaning the chromosphere consists of only one linear section. 
   Furthermore, in each of these active models the whole chromosphere is shifted inward about $\Delta m = 0.5\,$dex 
   compared to the inactive models. 
   The width of the chromosphere nearly remains the same in every case, but the temperature at the top of the 
   chromosphere increases at least by about $300\,$K. 
   The interaction of the free parameters yields chromospheric emission lines in the corresponding synthetic spectra, 
   i.\,e., varying only one parameter does not necessarily yield all the chromospheric lines in emission. 
   In particular, the contribution function of PHOENIX can provide information about where the lines are formed, 
   as discussed in Sect.~\ref{sec:contri}. 
   
   In these three active models all three investigated lines clearly appear in emission. 
   In Fig.~\ref{model_lc2_active}, we show 
   the observed spectrum of \mbox{TYC~3529-1437-1} with median activity as given by $I_\mathrm{Ca\, IRT}$, 
   the spectra of the inactive and active model, and the best linear-combination spectral fit of the two models to the observed spectrum. 
   From the modified $\chi^2_m$ and in comparison to Fig.~\ref{inactive_obs_mods_2} it becomes obvious that the combination 
   provides a better representation of the observation than the best-fit single model. 
   In Fig.~\ref{model_lc2_active} we also show the linear-combination fit for the spectrum of GJ~360. 
   Again the linear combination gives a better fit to the shape of the \ion{Na}{i}~D$_2$ and H$\alpha$ lines. However, reproducing 
   the transition from absorption to emission in the \ion{Ca}{ii}~IRT line also turns out to be hard with this method. 
   While in the fit for \mbox{TYC~3529-1437-1} the inactive model spectrum contributes $75\%$,
   the filling factors for the moderately more active \mbox{LP~733-099} yields $68\%$ for the inactive and $32\%$ for the active component. 
   The semi-active stars \mbox{G~234-057} and GJ~360 exhibit filling factors of $90\%$ and $82\%$ for the inactive chromospheric model. 
   We therefore conclude that the resulting filling factors of the active model component 
   match the activity levels of the stars as determined from spectral indices.

   \subsection{Filling factor as a function of activity level} \label{sec:ff_time_series}
   To study the relationship between the activity state and the filling factors in individual stars, 
   we perform an analysis of a set of available CARMENES spectra, 
   considering the restrictions described in Sect.~\ref{data_reduction}, of a subsample of stars. 
   This subsample consists of all active stars and those 
   inactive stars from which we have used more than five spectra 
   and that show variability in the flux density of the considered chromospheric lines. 
   We determine filling factors based on fits to all used CARMENES spectra, 
   following the same method as in Sect.~\ref{lin_combs}, but fix
   the combination of models to the pair previously determined.  
   
\begin{figure}[t]
   \centering
   \includegraphics[width=0.5\textwidth]{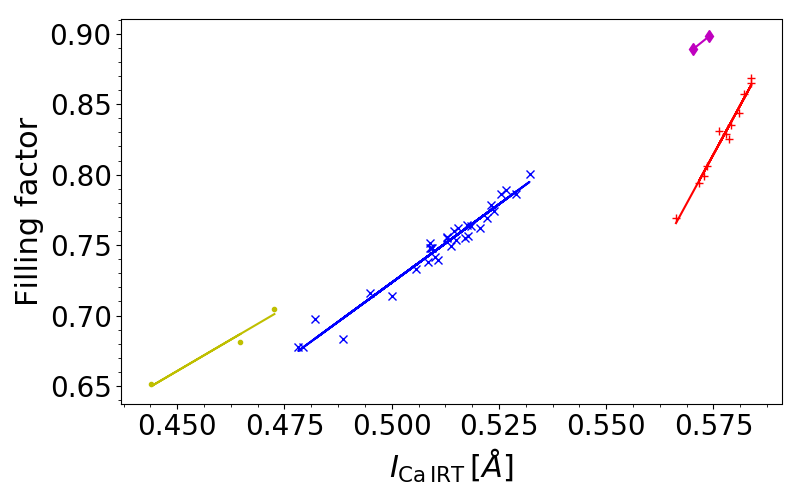}
   \includegraphics[width=0.5\textwidth]{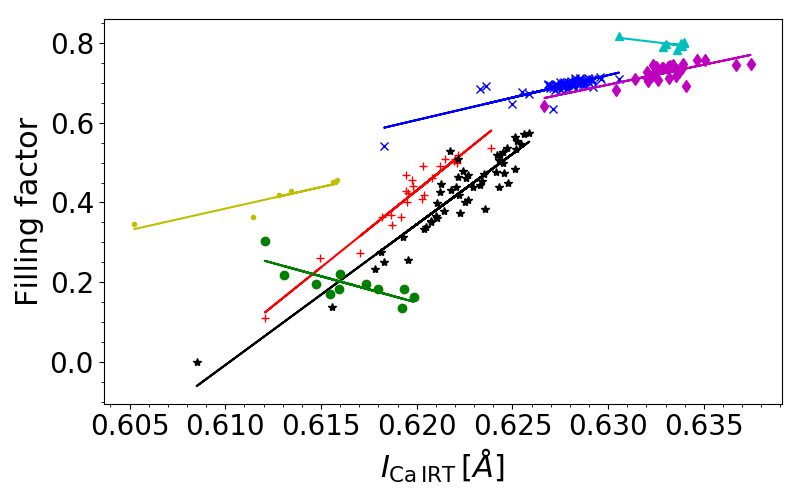}
      \caption{
      \textit{Upper panel:} Filling factors of the inactive models in the combination fits 
      as a function of $I_\mathrm{Ca\, IRT}$ of the four active stars \mbox{LP~733-099} (yellow dots), 
      \mbox{TYC~3529-1437-1} (blue crosses), GJ~360 (red pluses), and \mbox{G~234-057} (magenta diamonds) 
      are shown. 
      \textit{Lower panel:} Same as in the upper panel for Ross~905 (blue crosses), EW~Dra (red pluses), \mbox{LP~743-031} (yellow dots), 
      GJ~625 (magenta diamonds), BD+70~68 (black asterisks), Ross~730 (cyan triangles), and GJ~793 (green circles). 
      The solid lines show the linear fits of the stars. 
      To improve clarity, the errors of $I_\mathrm{Ca\, IRT}$ are not shown. 
              }
         \label{FigStellarSample_filling_factors}
   \end{figure}

   In Fig.~\ref{FigStellarSample_filling_factors} 
   we show the filling factor of the inactive model component as listed in 
   Table~\ref{table_stars_models_best_combined_8535} as a function of the $I_\mathrm{Ca\, IRT}$.  
   Our modeling yields a decrease in the filling factor of the inactive chromospheric component as the level of activity rises
   (i.e., $I_\mathrm{Ca\, IRT}$ decreases), but
   even for \mbox{LP~733-099}, the most active star in our sample, the filling factor of the inactive
   model component remains as high as $\sim 65\%$. 
   While this indicates that the major fraction of its surface
   is not covered with active chromosphere, we caution that our two-component
   approach remains a highly simplified description of the chromosphere. 
   In the case of the inactive stars, the interpretation of the filling factor in
   Table~\ref{table_stars_models_best_combined_8535} is more complicated. 
   In particular, combinations of two inactive chromospheric components possibly yield formally large
   filling factors for either component. 
   
   In the upper panel of Fig.~\ref{FigStellarSample_filling_factors} we show the relation between the
   Ca~IRT index, $I_\mathrm{Ca\, IRT}$, and the filling factor for
   the active stars \mbox{LP~733-099}, \mbox{TYC~3529-1437-1}, GJ~360, \mbox{G~234-057}. 
   There is a clear relation between $I_\mathrm{Ca\, IRT}$ and the filling factor in all these stars, which
   can well be approximated by a linear trend (see Fig.~\ref{FigStellarSample_filling_factors}). 
   This trend can also be seen for most of the inactive stars considered in this study, 
   some of these are plotted in the lower panel of Fig.~\ref{FigStellarSample_filling_factors}. 
   The filling factor of the inactive model decreases with increasing activity level. 
   
   Therefore linear regressions are performed to reveal the following linear trends:
   \begin{equation}\label{lin_reg}
   FF_{\rm star} = a  + b  \left( I_\mathrm{Ca\, IRT} - \overline{I_\mathrm{Ca\, IRT}} \right)
   \end{equation}
   for the filling factor ($FF_{\rm star}$) as a function of $I_\mathrm{Ca\, IRT}$, 
   where $\overline{I_\mathrm{Ca\, IRT}}$ is the mean $I_\mathrm{Ca\, IRT}$ of the respective star. 
   The gradients are denoted by $b$ and the intercepts by $a$, and the uncertainties on the coefficients ($\sigma_b$ and $\sigma_a$)
   were estimated using the Jackknife method \citep{Efron10.2307/2240822}. 
   For example, the application of a linear regression for GJ~360 yields $a = 0.83 \pm 0.0015$ and $b = 5.59 \pm 0.28 \, \AA^{-1}$. 
   Table~\ref{table_gradients} lists the results of the linear regressions of our study. 
   For nearly all of the subsample stars we obtain clear positive gradients. 
   Ross~730 and Wolf~1014 show a slight negative trend that cannot be distinguished from noise 
   as visible in the comparably high relative errors. 
   The negative slope of Ross~730 is dominated by a single outlier data point. 
   Therefore GJ~793 is the only star for which we find a significant negative gradient. 
   In this case model \#080 is more active than model \#112 according to $I_\mathrm{Ca\, IRT}$, although it is the opposite for H$\alpha$. 
   The variation of H$\alpha$ in GJ~793 leads the linear-combination method to give model \#112 a higher weight 
   with increasing activity state of GJ~793. 
   The particular model combination is the reason for the negative slope of GJ~793. 
   
   Interestingly, the range of filling factors of the inactive part for the active stars shown in Fig.~\ref{FigStellarSample_filling_factors}
   is comparable irrespective of their activity level. 
   However, with increasing $I_\mathrm{Ca\, IRT}$ the gradient of the relationship appears to increase 
   except for \mbox{G~234-057} from which we only used two spectra. 
   Therefore, compared to an active star, the same change in filling factor in an inactive star produces
   only a comparably weak response in the Ca~IRT line index, $I_\mathrm{Ca\, IRT}$. 
   In our modeling, this results from a stronger contrast between inactive and active chromospheric components in active stars, 
   which is also indicated by the differences of the \ion{Ca}{ii}~IRT line indices as given in Table~\ref{table_stars_models_best_combined_8535}.

\section{Conclusions}
   We have calculated a set of one-dimensional parametrized chromosphere models for a stellar sample 
   of M-type stars in the effective temperature range $3500\pm50\,$K. 
   The synthetic spectra of single models have turned out to be able to represent inactive stars 
   in the lines of \ion{Na}{i}~D$_2$, H$\alpha$, and a \ion{Ca}{ii}~IRT line simultaneously, 
   while a linear combination of at least two models is needed to simultaneously approach the chromospheric lines of 
   active stars, suggesting that the enhanced activity originates only in parts of the stellar surface -- as though the Sun 
   is covered partially in active regions. 
   The shape of the temperature structure of the models representing the inactive stars is comparable to the VAL~C model for the Sun. 
   A steep temperature gradient in the lower chromosphere is followed by a plateau-like structure in the upper chromosphere. 
   Our best-fit inactive models resemble this general structure of the VAL~C model, 
   but the temperatures and column mass densities differ, of course. 
   
   Concerning the models representing the active regions of the four active stars 
   in our sample, the temperature structure rather indicates 
   a steeper temperature gradient in the upper chromosphere than a plateau-like structure. 
   The deduced filling factors of inactive and active models correspond to the activity levels of the active stars under the constraint 
   of the model combination. 
   Furthermore, the 
   variable stars revealed a linear relationship between the filling factors and the line index in the 
   \ion{Ca}{ii}~IRT line, i.\,e., the higher the activity state, the less the coverage of the inactive chromosphere. 
   The gradients of the filling factors 
   of the variable stars depend on the model combinations, hence the gradients are not 
   evenly distributed, but they only vary in their absolute value. 
   Moreover, the model combination analysis also indicates an increasing contrast between the inactive and active regions with 
   increasing level of activity.

\begin{acknowledgements}
   D.H. acknowledges funding by the DLR under DLR 50 OR1701. B.F. 
   acknowledges funding by the DFG under Cz \mbox{222/1-1} and Schm 
   1032/69-1. 
   S.C. acknowledges support through DFG projects SCH 1382/2-1 and SCHM 1032/66-1.
   CARMENES is an instrument for the Centro Astron\'omico Hispano-Alem\'an 
   de
   Calar Alto (CAHA, Almer\'{\i}a, Spain).
   CARMENES is funded by the German Max-Planck-Gesellschaft (MPG), 
   the Spanish Consejo Superior de Investigaciones Cient\'{\i}ficas (CSIC),
   the European Union through FEDER/ERF FICTS-2011-02 funds,
   and the members of the CARMENES Consortium
   (Max-Planck-Institut f\"ur Astronomie, 
   Instituto de Astrof\'{\i}sica de Andaluc\'{\i}a, 
   Landessternwarte K\"onigstuhl, 
   Institut de Ci\`encies de l'Espai, 
   Institut f\"ur Astrophysik G\"ottingen, 
   Universidad Complutense de Madrid, 
   Th\"uringer Landessternwarte Tautenburg, 
   Instituto de Astrof\'{\i}sica de Canarias, 
   Hamburger Sternwarte, 
   Centro de Astrobiolog\'{\i}a and
   Centro Astron\'omico Hispano-Alem\'an), 
   with additional contributions by the Spanish Ministry of 
   Science  [through  projects  AYA2016-79425-C3-1/2/3-P, 
   ESP2016-80435-C2-1-R, AYA2015-69350-C3-2-P, and AYA2018-84089],
   the German Science Foundation through the Major Research Instrumentation
   Programme and DFG Research Unit FOR2544 ``Blue Planets around Red 
   Stars'',
   the Klaus Tschira Stiftung, 
   the states of Baden-W\"urttemberg and Niedersachsen, 
   and by the Junta de Andaluc\'{\i}a. 
   
   We thank J.~M.~Fontenla for providing us with the atmospheric structure of his 
   chromospheric model for GJ~832 \citep{Fontenla2016ApJ...830..154F} for comparison purposes. 
   
   CHIANTI is a collaborative project involving George Mason University, the University of Michigan (USA) and the University of Cambridge (UK). 
\end{acknowledgements}

\bibliographystyle{aa} 
\bibliography{references}

\begin{appendix} 

\section{Best single-component fits and best linear-combination fits}

   \begin{table}[h!]
   \caption{Best single-component fits for the considered stars.}             
   \label{table_stars_models_best_mean}      
   \centering
   \begin{tabular}{l c c}
   \hline\hline       
Stars   & Model   & $\chi^2_m$ \\ 
 \hline 
Wolf 1056   & \#047   & 2.75 \\ 
GJ 47   & \#047   & 2.53 \\ 
BD+70 68   & \#080   & 2.26 \\ 
GJ 70   & \#079   & 2.29 \\ 
G 244-047   & \#042   & 3.45 \\ 
VX Ari   & \#079   & 3.61 \\ 
Ross 567   & \#042   & 2.75 \\ 
GJ 226   & \#047   & 1.8 \\ 
GJ 258   & \#047   & 3.83 \\ 
GJ 1097   & \#042   & 4.04 \\ 
GJ 3452   & \#042   & 2.37 \\ 
G 234-057$^\bigstar$   & \#136   & 10.17 \\ 
GJ 357   & \#042   & 2.63 \\ 
GJ 360$^\bigstar$   & \#136   & 10.12 \\ 
GJ 386   & \#047   & 2.78 \\ 
LP 670-017   & \#080   & 4.51 \\ 
GJ 399   & \#079   & 2.54 \\ 
Ross 104   & \#079   & 2.59 \\ 
LP 733-099$^\bigstar$   & \#131   & 28.11 \\ 
Ross 905   & \#042   & 2.84 \\ 
GJ 443   & \#080   & 2.04 \\ 
Ross 690   & \#079   & 2.41 \\ 
Ross 695   & \#042   & 2.91 \\ 
Ross 992   & \#079   & 2.85 \\ 
$\theta$ Boo B   & \#047   & 2.55 \\ 
Ross 1047   & \#047   & 3.16 \\ 
LP 743-031   & \#080   & 3.38 \\ 
G 137-084   & \#080   & 2.66 \\ 
EW Dra   & \#080   & 2.31 \\ 
GJ 625   & \#042   & 2.41 \\ 
GJ 1203   & \#047   & 2.5 \\ 
LP 446-006   & \#047   & 2.82 \\ 
Ross 863   & \#079   & 3.11 \\ 
GJ 2128   & \#042   & 2.45 \\ 
GJ 671   & \#042   & 3.3 \\ 
G 204-039   & \#080   & 2.79 \\ 
TYC 3529-1437-1$^\bigstar$   & \#131   & 18.2 \\ 
Ross 145   & \#042   & 3.28 \\ 
G 155-042   & \#042   & 3.54 \\ 
Ross 730   & \#029   & 2.82 \\ 
HD 349726   & \#029   & 2.83 \\ 
GJ 793   & \#080   & 3.01 \\ 
Wolf 896   & \#047   & 2.59 \\ 
Wolf 906   & \#079   & 2.43 \\ 
LSPM J2116+0234   & \#079   & 3.08 \\ 
Wolf 926   & \#047   & 2.99 \\ 
BD-05 5715   & \#080   & 2.9 \\ 
Wolf 1014   & \#042   & 3.32 \\ 
G 273-093   & \#047   & 1.97 \\ 
Wolf 1051   & \#080   & 2.31 \\ 
   \hline   
   \end{tabular}
   \tablefoot{
   Asterisks identify active stars in the stellar sample. 
   Figure~\ref{model_grid_inactive_active} illustrates the temperature structure of all the best single-component fits. 
   The spectra of the best-fit models for GJ~671 and EW~Dra are shown in Fig.~\ref{inactive_obs_mods}, 
   and those for GJ~360 and \mbox{TYC~3529-1437-1} are shown in Fig.~\ref{inactive_obs_mods_2}. 
   }
   \end{table}

   \begin{table*}[t!]
   \caption{Best-fit models in a linear-combination fit with filling factors and the difference $\Delta I_\mathrm{Ca\, IRT}$ of the models.}             
   \label{table_stars_models_best_combined_8535}      
   \centering
   \begin{tabular}{l c c c c c c}
   \hline\hline       
Stars   & \multicolumn{2}{c}{Inactive model} & \multicolumn{2}{c}{More active model} & $\chi^2_m$ & $\Delta I_\mathrm{Ca\, IRT}$ \\ 
        & Model & Filling factor & Model & Filling factor &  & $[\AA]$ \\ 
 \hline 
Wolf 1056   & \#047   & 0.8   & \#050   & 0.2   & 1.02   & 0.047 \\ 
GJ 47   & \#047   & 0.81   & \#050   & 0.19   & 0.86   & 0.047 \\ 
BD+70 68   & \#047   & 0.4   & \#080   & 0.6   & 1.18   & 0.033 \\ 
GJ 70   & \#079   & 0.72   & \#049   & 0.28   & 1.18   & 0.067 \\ 
G 244-047   & \#042   & 0.66   & \#049   & 0.34   & 1.08   & 0.071 \\ 
VX Ari   & \#079   & 0.66   & \#049   & 0.34   & 2.04   & 0.067 \\ 
Ross 567   & \#042   & 0.73   & \#049   & 0.27   & 1.25   & 0.071 \\ 
GJ 226   & \#047   & 0.92   & \#064   & 0.08   & 0.66   & 0.189 \\ 
GJ 258   & \#047   & 0.87   & \#064   & 0.13   & 1.08   & 0.189 \\ 
GJ 1097   & \#042   & 0.63   & \#049   & 0.37   & 1.29   & 0.071 \\ 
GJ 3452   & \#042   & 0.76   & \#049   & 0.24   & 1.22   & 0.071 \\ 
G 234-057$^\bigstar$   & \#080   & 0.9   & \#139   & 0.1   & 3.58   & 0.389 \\ 
GJ 357   & \#042   & 0.74   & \#049   & 0.26   & 1.24   & 0.071 \\ 
GJ 360$^\bigstar$   & \#080   & 0.82   & \#132   & 0.17   & 5.13   & 0.215 \\ 
GJ 386   & \#047   & 0.81   & \#050   & 0.19   & 1.21   & 0.047 \\ 
LP 670-017   & \#042   & 0.49   & \#080   & 0.51   & 2.23   & 0.039 \\ 
GJ 399   & \#079   & 0.8   & \#065   & 0.2   & 1.31   & 0.093 \\ 
Ross 104   & \#079   & 0.71   & \#049   & 0.29   & 1.46   & 0.067 \\ 
LP 733-099$^\bigstar$   & \#079   & 0.68   & \#149   & 0.32   & 16.13   & 0.598 \\ 
Ross 905   & \#042   & 0.7   & \#049   & 0.3   & 1.05   & 0.071 \\ 
GJ 443   & \#047   & 0.37   & \#080   & 0.63   & 1.09   & 0.033 \\ 
Ross 690   & \#079   & 0.82   & \#065   & 0.18   & 1.39   & 0.093 \\ 
Ross 695   & \#042   & 0.73   & \#049   & 0.27   & 1.41   & 0.071 \\ 
Ross 992   & \#079   & 0.68   & \#049   & 0.32   & 1.43   & 0.067 \\ 
$\theta$ Boo B   & \#047   & 0.8   & \#050   & 0.2   & 0.83   & 0.047 \\ 
Ross 1047   & \#047   & 0.88   & \#064   & 0.12   & 1.01   & 0.189 \\ 
LP 743-031   & \#047   & 0.43   & \#080   & 0.57   & 2.12   & 0.033 \\ 
G 137-084   & \#029   & 0.39   & \#080   & 0.61   & 1.39   & 0.038 \\ 
EW Dra   & \#047   & 0.44   & \#080   & 0.56   & 0.96   & 0.033 \\ 
GJ 625   & \#042   & 0.74   & \#049   & 0.26   & 1.02   & 0.071 \\ 
GJ 1203   & \#047   & 0.81   & \#050   & 0.19   & 0.98   & 0.047 \\ 
LP 446-006   & \#047   & 0.8   & \#050   & 0.2   & 1.08   & 0.047 \\ 
Ross 863   & \#079   & 0.8   & \#060   & 0.2   & 1.87   & 0.102 \\ 
GJ 2128   & \#042   & 0.75   & \#049   & 0.25   & 1.13   & 0.071 \\ 
GJ 671   & \#042   & 0.7   & \#049   & 0.3   & 1.44   & 0.071 \\ 
G 204-039   & \#029   & 0.36   & \#080   & 0.64   & 1.73   & 0.038 \\ 
TYC 3529-1437-1$^\bigstar$   & \#079   & 0.75   & \#149   & 0.25   & 14.25   & 0.598 \\ 
Ross 145   & \#042   & 0.69   & \#049   & 0.31   & 1.27   & 0.071 \\ 
G 155-042   & \#042   & 0.68   & \#049   & 0.32   & 1.44   & 0.071 \\ 
Ross 730   & \#029   & 0.8   & \#049   & 0.2   & 1.89   & 0.070 \\ 
HD 349726   & \#029   & 0.8   & \#049   & 0.2   & 1.91   & 0.070 \\ 
GJ 793   & \#112   & 0.22   & \#080   & 0.78   & 1.93   & 0.050 \\ 
Wolf 896   & \#047   & 0.89   & \#064   & 0.11   & 0.67   & 0.189 \\ 
Wolf 906   & \#079   & 0.79   & \#065   & 0.21   & 1.09   & 0.093 \\ 
LSPM J2116+0234   & \#079   & 0.78   & \#060   & 0.22   & 1.54   & 0.102 \\ 
Wolf 926   & \#047   & 0.79   & \#050   & 0.21   & 1.05   & 0.047 \\ 
BD-05 5715   & \#047   & 0.43   & \#080   & 0.57   & 1.66   & 0.033 \\ 
Wolf 1014   & \#042   & 0.7   & \#049   & 0.3   & 1.43   & 0.071 \\ 
G 273-093   & \#047   & 0.85   & \#050   & 0.15   & 0.92   & 0.047 \\ 
Wolf 1051   & \#042   & 0.32   & \#080   & 0.68   & 1.44   & 0.039 \\ 
   \hline   
   \end{tabular}               
   \tablefoot{
   The line of \ion{Na}{i}~$\mathrm{D}_2$ and the bluest \ion{Ca}{ii}~IRT line 
   are weighted by a factor of $12$ compared to H$\alpha$ as described in Sect.~\ref{sec:comparison}. 
   Asterisks identify active stars in the stellar sample. 
   The combinations for GJ~360 and \mbox{TYC~3529-1437-1} are illustrated in Fig.~\ref{model_lc2_active}. 
   }
   \end{table*}
   
\vspace*{24cm}

\newpage

\vspace*{24cm}
\newpage

\section{Linear regressions of the filling factors as a function of activity state}
\vspace*{4cm}

   \begin{table}[h!]
   \caption{Gradients $b \pm \sigma_b$ and intercepts $a \pm \sigma_a$ of the linear regressions 
            of the filling factors of the inactive region as a function of $I_\mathrm{Ca\, IRT}$ 
            (Eq.~\ref{lin_reg}). 
            }             
   \label{table_gradients}      
   \centering
   \begin{tabular}{l c c c c}
   \hline\hline       
Stars   & $b \, [\AA^{-1}]$ & $a $ \\ 
 \hline   
Wolf 1056   & 1.33 $\pm$ 19.94   & 0.79 $\pm$ 0.0128 \\ 
BD+70 68   & 35.28 $\pm$ 3.60   & 0.42 $\pm$ 0.0060 \\ 
G 244-047   & 9.66 $\pm$ 6.31   & 0.63 $\pm$ 0.0056 \\ 
Ross 567   & 5.01 $\pm$ 2.13   & 0.71 $\pm$ 0.0033 \\ 
GJ 258   & 2.27 $\pm$ 2.60   & 0.87 $\pm$ 0.0034 \\ 
GJ 1097   & 10.96 $\pm$ 11.49   & 0.61 $\pm$ 0.0169 \\ 
G 234-057$^\bigstar$   & 2.44 $\pm$ 239.72   & 0.89 $\pm$ 0.4470 \\ 
GJ 360$^\bigstar$   & 5.59 $\pm$ 0.28   & 0.83 $\pm$ 0.0015 \\ 
Ross 104   & 5.86 $\pm$ 2.31   & 0.72 $\pm$ 0.0017 \\ 
LP 733-099$^\bigstar$   & 1.78 $\pm$ 0.98   & 0.68 $\pm$ 0.0092 \\ 
Ross 905   & 11.25 $\pm$ 4.82   & 0.69 $\pm$ 0.0019 \\ 
Ross 690   & 4.13 $\pm$ 0.96   & 0.81 $\pm$ 0.0012 \\ 
Ross 992   & 4.77 $\pm$ 3.19   & 0.69 $\pm$ 0.0056 \\ 
Ross 1047   & 7.95 $\pm$ 1.35   & 0.87 $\pm$ 0.0036 \\ 
LP 743-031   & 10.80 $\pm$ 7.01   & 0.41 $\pm$ 0.0143 \\ 
G 137-084   & 31.95 $\pm$ 4.38   & 0.39 $\pm$ 0.0053 \\ 
EW Dra   & 38.59 $\pm$ 2.54   & 0.42 $\pm$ 0.0060 \\ 
GJ 625   & 10.03 $\pm$ 2.93   & 0.73 $\pm$ 0.0034 \\ 
LP 446-006   & 2.67 $\pm$ 7.99   & 0.81 $\pm$ 0.0107 \\ 
Ross 863   & 5.51 $\pm$ 4.24   & 0.80 $\pm$ 0.0063 \\ 
GJ 2128   & 17.22 $\pm$ 4.75   & 0.73 $\pm$ 0.0052 \\ 
GJ 671   & 14.65 $\pm$ 3.31   & 0.68 $\pm$ 0.0028 \\ 
G 204-039   & 29.84 $\pm$ 6.28   & 0.29 $\pm$ 0.0097 \\ 
TYC 3529-1437-1$^\bigstar$   & 2.23 $\pm$ 0.11   & 0.75 $\pm$ 0.0011 \\ 
Ross 145   & 12.94 $\pm$ 2.05   & 0.67 $\pm$ 0.0052 \\ 
Ross 730   & -5.76 $\pm$ 10.77   & 0.80 $\pm$ 0.0055 \\ 
HD 349726   & 0.31 $\pm$ 7.66   & 0.78 $\pm$ 0.0048 \\ 
GJ 793   & -13.19 $\pm$ 5.83   & 0.20 $\pm$ 0.0100 \\ 
Wolf 896   & 6.37 $\pm$ 0.69   & 0.88 $\pm$ 0.0024 \\ 
LSPM J2116+0234   & 7.07 $\pm$ 1.06   & 0.78 $\pm$ 0.0011 \\ 
Wolf 926   & 13.78 $\pm$ 2.52   & 0.77 $\pm$ 0.0042 \\ 
BD-05 5715   & 19.60 $\pm$ 6.67   & 0.44 $\pm$ 0.0080 \\ 
Wolf 1014   & -1.07 $\pm$ 2.63   & 0.69 $\pm$ 0.0030 \\ 
   \hline   
   \end{tabular}
   \tablefoot{
   Asterisks identify active stars in the stellar sample.
   }
   \end{table}

\section{Calculated model set}

\longtab[1]{
\begin{longtable}{c c c c c c c c}
\caption{\label{table_model_grid} Calculated model set and parameters.}\\
\hline\hline
Model & $m_\mathrm{min}$ & $m_\mathrm{mid}$ & $T_\mathrm{mid}$ & $m_\mathrm{top}$ & $T_\mathrm{top}$ & $grad_{\mathrm{TR}}$          & $I_\mathrm{Ca\, IRT}$ \\ 
      & $[\mathrm{dex}]$ & $[\mathrm{dex}]$ & $[\mathrm{K}]$   & $[\mathrm{dex}]$ & $[\mathrm{K}]$   & $[\mathrm{dex}]$ & $[\AA]$ \\ 
\hline
\endfirsthead
\caption{Continued.}\\
\hline\hline     
Model & $m_\mathrm{min}$ & $m_\mathrm{mid}$ & $T_\mathrm{mid}$ & $m_\mathrm{top}$ & $T_\mathrm{top}$ & $grad_{\mathrm{TR}}$          & $I_\mathrm{Ca\, IRT}$ \\ 
      & $[\mathrm{dex}]$ & $[\mathrm{dex}]$ & $[\mathrm{K}]$   & $[\mathrm{dex}]$ & $[\mathrm{K}]$   & $[\mathrm{dex}]$ & $[\AA]$ \\ 
\hline
\endhead
\hline  
\endfoot

\#001   & -4.0   & -4.3   & 5500   & -6.0   & 6000   & 7.5   & 0.653 \\ 
\#002   & -3.5   & -3.6   & 4500   & -5.2   & 5000   & 7.5   & 0.655 \\ 
\#003   & -3.5   & -3.6   & 4500   & -5.2   & 5000   & 8.0   & 0.654 \\ 
\#004   & -3.5   & -3.6   & 4500   & -5.1   & 5000   & 7.5   & 0.655 \\ 
\#005   & -3.5   & -3.8   & 5500   & -5.5   & 6000   & 7.5   & 0.655 \\ 
\#006   & -3.4   & -3.6   & 4500   & -5.1   & 5000   & 7.5   & 0.654 \\ 
\#007   & -3.2   & -3.6   & 4500   & -5.1   & 5000   & 7.5   & 0.653 \\ 
\#008   & -3.2   & -3.6   & 4500   & -5.0   & 5000   & 7.5   & 0.654 \\ 
\#009   & -3.2   & -3.6   & 5500   & -5.1   & 6000   & 7.5   & 0.655 \\ 
\#010   & -3.2   & -3.5   & 4500   & -5.0   & 5000   & 7.5   & 0.654 \\ 
\#011   & -3.2   & -3.5   & 4700   & -5.0   & 5200   & 7.5   & 0.654 \\ 
\#012   & -3.1   & -3.6   & 4500   & -5.0   & 5000   & 8.5   & 0.652 \\ 
\#013   & -3.1   & -3.3   & 4500   & -5.0   & 5000   & 8.5   & 0.653 \\ 
\#014   & -3.1   & -3.3   & 5500   & -5.0   & 6000   & 8.5   & 0.654 \\ 
\#015   & -3.0   & -3.6   & 4500   & -5.0   & 5000   & 7.5   & 0.653 \\ 
\#016   & -3.0   & -3.6   & 5500   & -5.0   & 6000   & 7.5   & 0.652 \\ 
\#017   & -3.0   & -3.4   & 4500   & -4.9   & 5000   & 7.5   & 0.653 \\ 
\#018   & -3.0   & -3.3   & 4500   & -5.0   & 5000   & 7.5   & 0.653 \\ 
\#019   & -3.0   & -3.3   & 4500   & -5.0   & 5000   & 7.5   & 0.653 \\ 
\#020   & -3.0   & -3.3   & 5500   & -5.0   & 6000   & 7.5   & 0.653 \\ 
\#021   & -3.0   & -3.3   & 5500   & -5.0   & 6000   & 7.5   & 0.653 \\ 
\#022   & -2.8   & -3.6   & 4500   & -5.0   & 5000   & 7.5   & 0.576 \\ 
\#023   & -2.8   & -3.6   & 5500   & -5.0   & 6000   & 7.5   & 0.651 \\ 
\#024   & -2.8   & -3.3   & 4500   & -5.0   & 5000   & 7.5   & 0.653 \\ 
\#025   & -2.8   & -3.3   & 5500   & -5.0   & 6000   & 7.5   & 0.651 \\ 
\#026   & -2.6   & -3.6   & 4500   & -5.0   & 5000   & 8.5   & 0.650 \\ 
\#027   & -2.6   & -3.2   & 4500   & -4.5   & 5000   & 8.5   & 0.652 \\ 
\#028   & -2.6   & -3.2   & 4500   & -4.5   & 6000   & 8.5   & 0.650 \\ 
\#029   & -2.6   & -3.2   & 4500   & -4.5   & 7000   & 8.5   & 0.649 \\ 
\#030   & -2.6   & -3.0   & 4500   & -4.5   & 5000   & 8.5   & 0.652 \\ 
\#031   & -2.6   & -3.0   & 4500   & -4.5   & 6000   & 8.5   & 0.651 \\ 
\#032   & -2.6   & -3.0   & 4500   & -4.5   & 7000   & 8.5   & 0.651 \\ 
\#033   & -2.6   & -2.8   & 3500   & -4.5   & 5000   & 8.5   & 0.653 \\ 
\#034   & -2.6   & -2.8   & 4000   & -4.5   & 4500   & 8.5   & 0.652 \\ 
\#035   & -2.6   & -2.8   & 4500   & -4.5   & 5000   & 8.0   & 0.653 \\ 
\#036   & -2.6   & -2.8   & 4500   & -4.5   & 5000   & 8.5   & 0.653 \\ 
\#037   & -2.6   & -2.8   & 4500   & -4.5   & 6000   & 8.5   & 0.652 \\ 
\#038   & -2.6   & -2.8   & 4500   & -4.5   & 7000   & 8.5   & 0.652 \\ 
\#039   & -2.5   & -3.6   & 4500   & -5.0   & 5000   & 7.5   & 0.652 \\ 
\#040   & -2.5   & -3.6   & 5500   & -5.0   & 6000   & 7.5   & 0.650 \\ 
\#041   & -2.5   & -3.6   & 5500   & -5.0   & 6200   & 7.5   & 0.650 \\ 
\#042   & -2.5   & -3.6   & 5500   & -5.0   & 6500   & 7.5   & 0.650 \\ 
\#043   & -2.5   & -3.3   & 4500   & -5.0   & 5000   & 7.5   & 0.652 \\ 
\#044   & -2.5   & -3.3   & 5500   & -5.0   & 6000   & 7.5   & 0.651 \\ 
\#045   & -2.5   & -2.8   & 4500   & -4.5   & 5000   & 7.5   & 0.654 \\ 
\#046   & -2.5   & -2.8   & 5500   & -4.5   & 6000   & 7.5   & 0.649 \\ 
\#047   & -2.5   & -2.7   & 6500   & -5.0   & 7000   & 9.2   & 0.644 \\ 
\#048   & -2.1   & -2.6   & 4500   & -4.0   & 5000   & 8.5   & 0.658 \\ 
\#049   & -2.1   & -2.6   & 6500   & -4.5   & 7000   & 7.5   & 0.579 \\ 
\#050   & -2.1   & -2.6   & 6500   & -4.0   & 7000   & 9.2   & 0.597 \\ 
\#051   & -2.1   & -2.3   & 4500   & -4.0   & 5000   & 8.5   & 0.664 \\ 
\#052   & -2.1   & -2.3   & 5000   & -5.0   & 5500   & 7.5   & 0.667 \\ 
\#053   & -2.1   & -2.3   & 5000   & -4.0   & 6000   & 8.5   & 0.648 \\ 
\#054   & -2.1   & -2.3   & 5000   & -4.0   & 6000   & 9.0   & 0.660 \\ 
\#055   & -2.1   & -2.3   & 5000   & -4.0   & 6000   & 9.5   & 0.665 \\ 
\#056   & -2.1   & -2.3   & 5500   & -5.0   & 6000   & 7.5   & 0.660 \\ 
\#057   & -2.1   & -2.3   & 5500   & -4.5   & 6000   & 7.5   & 0.632 \\ 
\#058   & -2.1   & -2.3   & 5500   & -4.0   & 6000   & 8.5   & 0.628 \\ 
\#059   & -2.1   & -2.3   & 5500   & -4.0   & 6000   & 9.5   & 0.649 \\ 
\#060   & -2.1   & -2.3   & 6500   & -5.0   & 7000   & 9.2   & 0.543 \\ 
\#061   & -2.1   & -2.3   & 6500   & -4.5   & 7000   & 9.2   & 0.524 \\ 
\#062   & -2.1   & -2.3   & 6500   & -4.2   & 7000   & 8.2   & 0.427 \\ 
\#063   & -2.1   & -2.3   & 6500   & -4.0   & 7000   & 9.0   & 0.442 \\ 
\#064   & -2.1   & -2.3   & 6500   & -4.0   & 7000   & 9.2   & 0.455 \\ 
\#065   & -2.0   & -2.5   & 6500   & -5.0   & 8000   & 9.2   & 0.552 \\ 
\#066   & -2.0   & -2.5   & 6500   & -4.5   & 8000   & 9.2   & 0.510 \\ 
\#067   & -2.0   & -2.3   & 5000   & -4.0   & 7000   & 8.2   & 0.605 \\ 
\#068   & -2.0   & -2.3   & 6500   & -5.0   & 7000   & 9.2   & 0.539 \\ 
\#069   & -2.0   & -2.3   & 6500   & -4.5   & 7000   & 9.2   & 0.521 \\ 
\#070   & -2.0   & -2.3   & 6500   & -4.0   & 7000   & 9.2   & 0.455 \\ 
\#071   & -2.0   & -2.2   & 6500   & -5.0   & 7000   & 9.2   & 0.539 \\ 
\#072   & -2.0   & -2.2   & 6500   & -4.5   & 7000   & 9.2   & 0.521 \\ 
\#073   & -1.9   & -2.3   & 6500   & -4.0   & 7000   & 9.2   & 0.428 \\ 
\#074   & -1.8   & -2.6   & 3000   & -4.0   & 7000   & 9.0   & 0.653 \\ 
\#075   & -1.8   & -2.6   & 3500   & -4.0   & 7000   & 9.0   & 0.656 \\ 
\#076   & -1.8   & -2.3   & 3000   & -4.0   & 7000   & 9.0   & 0.656 \\ 
\#077   & -1.8   & -2.3   & 3500   & -4.0   & 7000   & 9.0   & 0.658 \\ 
\#078   & -1.8   & -2.3   & 6500   & -4.0   & 7000   & 9.0   & 0.363 \\ 
\#079   & -1.5   & -2.5   & 5000   & -4.0   & 7500   & 8.5   & 0.645 \\ 
\#080   & -1.5   & -2.5   & 5500   & -4.0   & 7500   & 8.5   & 0.611 \\ 
\#081   & -1.5   & -2.5   & 6000   & -4.1   & 7500   & 8.5   & 0.538 \\ 
\#082   & -1.5   & -2.5   & 6000   & -4.0   & 7500   & 8.0   & 0.439 \\ 
\#083   & -1.5   & -2.5   & 6000   & -4.0   & 7500   & 8.5   & 0.527 \\ 
\#084   & -1.5   & -2.5   & 6000   & -4.0   & 8000   & 8.0   & 0.312 \\ 
\#085   & -1.5   & -2.5   & 6000   & -3.9   & 7500   & 8.5   & 0.455 \\ 
\#086   & -1.5   & -2.5   & 6000   & -3.8   & 7500   & 8.5   & 0.424 \\ 
\#087   & -1.5   & -2.5   & 6000   & -3.7   & 6500   & 8.5   & 0.416 \\ 
\#088   & -1.5   & -2.5   & 6000   & -3.7   & 7000   & 8.5   & 0.391 \\ 
\#089   & -1.5   & -2.5   & 6000   & -3.7   & 7500   & 8.5   & 0.270 \\ 
\#090   & -1.5   & -2.5   & 6500   & -4.0   & 7500   & 8.0   & 0.111 \\ 
\#091   & -1.5   & -2.5   & 6500   & -3.5   & 7500   & 8.5   & -0.326 \\ 
\#092   & -1.5   & -2.3   & 4000   & -4.0   & 7000   & 9.0   & 0.665 \\ 
\#093   & -1.5   & -2.3   & 5000   & -4.0   & 7000   & 8.2   & 0.604 \\ 
\#094   & -1.5   & -2.3   & 5000   & -4.0   & 7000   & 9.0   & 0.650 \\ 
\#095   & -1.5   & -2.3   & 6500   & -4.5   & 7000   & 9.0   & 0.408 \\ 
\#096   & -1.5   & -2.3   & 6500   & -4.5   & 8000   & 9.2   & 0.089 \\ 
\#097   & -1.5   & -2.3   & 6500   & -4.0   & 7000   & 9.0   & 0.302 \\ 
\#098   & -1.5   & -2.0   & 4500   & -3.5   & 5000   & 8.5   & 0.668 \\ 
\#099   & -1.5   & -2.0   & 5000   & -3.5   & 5500   & 8.0   & 0.584 \\ 
\#100   & -1.5   & -2.0   & 5000   & -3.5   & 5500   & 8.5   & 0.599 \\ 
\#101   & -1.5   & -2.0   & 5000   & -3.5   & 5700   & 8.5   & 0.562 \\ 
\#102   & -1.5   & -2.0   & 5000   & -3.5   & 5900   & 8.5   & 0.510 \\ 
\#103   & -1.5   & -2.0   & 5000   & -3.5   & 6000   & 8.5   & 0.487 \\ 
\#104   & -1.5   & -1.7   & 5500   & -4.0   & 6000   & 8.0   & 0.444 \\ 
\#105   & -1.5   & -1.7   & 5500   & -4.0   & 6000   & 8.5   & 0.537 \\ 
\#106   & -1.5   & -1.7   & 6000   & -4.0   & 6500   & 8.5   & 0.366 \\ 
\#107   & -1.0   & -3.5   & 4000   & -5.0   & 8000   & 9.5   & 0.663 \\ 
\#108   & -1.0   & -3.5   & 4000   & -4.5   & 8000   & 9.5   & 0.666 \\ 
\#109   & -1.0   & -3.0   & 4000   & -5.0   & 8000   & 9.5   & 0.659 \\ 
\#110   & -1.0   & -3.0   & 4000   & -4.5   & 8000   & 9.5   & 0.662 \\ 
\#111   & -1.0   & -3.0   & 4000   & -4.0   & 8000   & 9.5   & 0.672 \\ 
\#112   & -1.0   & -2.7   & 4000   & -3.5   & 7000   & 9.5   & 0.661 \\ 
\#113   & -1.0   & -2.7   & 4000   & -3.5   & 7500   & 9.5   & 0.650 \\ 
\#114   & -1.0   & -2.7   & 4000   & -3.5   & 8000   & 9.5   & 0.607 \\ 
\#115   & -1.0   & -2.7   & 4000   & -3.5   & 8100   & 9.5   & 0.580 \\ 
\#116   & -1.0   & -2.7   & 4000   & -3.5   & 8200   & 9.5   & 0.539 \\ 
\#117   & -1.0   & -2.7   & 4000   & -3.5   & 8250   & 9.5   & 0.514 \\ 
\#118   & -1.0   & -2.7   & 4200   & -3.5   & 8200   & 9.5   & 0.516 \\ 
\#119   & -1.0   & -2.7   & 4200   & -3.5   & 8250   & 9.5   & 0.484 \\ 
\#120   & -1.0   & -2.7   & 4500   & -3.5   & 8300   & 8.5   & 0.256 \\ 
\#121   & -1.0   & -2.7   & 4600   & -3.6   & 7800   & 8.5   & 0.468 \\ 
\#122   & -1.0   & -2.6   & 5000   & -3.6   & 7800   & 8.5   & 0.411 \\ 
\#123   & -1.0   & -2.6   & 5000   & -3.6   & 7800   & 9.5   & 0.599 \\ 
\#124   & -1.0   & -2.5   & 4000   & -4.0   & 8000   & 9.5   & 0.671 \\ 
\#125   & -1.0   & -2.5   & 4300   & -3.5   & 8250   & 8.5   & 0.190 \\ 
\#126   & -1.0   & -2.5   & 4600   & -3.6   & 7800   & 8.5   & 0.380 \\ 
\#127   & -1.0   & -2.5   & 5200   & -3.8   & 8000   & 8.5   & 0.503 \\ 
\#128   & -1.0   & -2.5   & 5200   & -3.8   & 8200   & 8.5   & 0.425 \\ 
\#129   & -1.0   & -2.5   & 5200   & -3.7   & 8000   & 8.5   & 0.345 \\ 
\#130   & -1.0   & -2.5   & 5200   & -3.7   & 8000   & 8.8   & 0.424 \\ 
\#131   & -1.0   & -2.5   & 5200   & -3.7   & 8000   & 9.5   & 0.534 \\ 
\#132   & -1.0   & -2.5   & 5200   & -3.6   & 8000   & 9.6   & 0.396 \\ 
\#133   & -1.0   & -2.5   & 5200   & -3.5   & 8000   & 9.5   & 0.324 \\ 
\#134   & -1.0   & -2.5   & 5500   & -3.5   & 8000   & 9.5   & 0.194 \\ 
\#135   & -1.0   & -2.3   & 4550   & -3.5   & 8000   & 8.5   & 0.155 \\ 
\#136   & -1.0   & -2.3   & 4600   & -3.8   & 7800   & 8.5   & 0.572 \\ 
\#137   & -1.0   & -2.3   & 4600   & -3.7   & 7800   & 8.5   & 0.452 \\ 
\#138   & -1.0   & -2.3   & 4600   & -3.6   & 7800   & 8.5   & 0.286 \\ 
\#139   & -1.0   & -2.3   & 4600   & -3.5   & 7800   & 8.5   & 0.222 \\ 
\#140   & -1.0   & -2.3   & 4600   & -3.5   & 7800   & 10.0   & 0.529 \\ 
\#141   & -1.0   & -2.3   & 4800   & -3.6   & 7800   & 8.5   & 0.240 \\ 
\#142   & -1.0   & -2.3   & 5000   & -3.7   & 7800   & 8.5   & 0.372 \\ 
\#143   & -1.0   & -2.3   & 5000   & -3.7   & 8000   & 8.5   & 0.285 \\ 
\#144   & -1.0   & -2.3   & 5000   & -3.6   & 7800   & 8.5   & 0.184 \\ 
\#145   & -1.0   & -2.3   & 5000   & -3.6   & 8000   & 8.5   & 0.073 \\ 
\#146   & -1.0   & -2.3   & 5000   & -3.5   & 7800   & 8.5   & 0.115 \\ 
\#147   & -1.0   & -2.3   & 5000   & -3.5   & 7900   & 8.5   & 0.060 \\ 
\#148   & -1.0   & -2.3   & 5500   & -3.5   & 8000   & 9.8   & 0.030 \\ 
\#149   & -1.0   & -2.3   & 5500   & -3.5   & 8000   & 10.0   & 0.047 \\ 
\#150   & -1.0   & -2.3   & 6000   & -3.5   & 8000   & 10.0   & -0.381 \\ 
\#151   & -1.0   & -2.2   & 5500   & -3.6   & 8000   & 9.5   & 0.063 \\ 
\#152   & -1.0   & -2.0   & 4000   & -4.0   & 8000   & 9.5   & 0.661 \\ 
\#153   & -1.0   & -2.0   & 4600   & -3.6   & 7800   & 8.5   & 0.193 \\ 
\#154   & -1.0   & -2.0   & 5200   & -3.5   & 8000   & 9.5   & -0.021 \\ 
\#155   & -1.0   & -1.5   & 4000   & -4.0   & 8000   & 9.5   & 0.576 \\ 
\#156   & -1.0   & -1.5   & 5000   & -3.5   & 6000   & 8.5   & 0.332 \\ 
\#157   & -0.5   & -2.7   & 4200   & -3.5   & 8200   & 9.5   & 0.515 \\ 
\#158   & -0.5   & -2.7   & 4200   & -3.5   & 8250   & 9.5   & 0.483 \\ 
\#159   & -0.3   & -3.0   & 4000   & -4.5   & 8000   & 9.5   & 0.675 \\ 
\#160   & -0.3   & -2.8   & 4000   & -3.5   & 7000   & 9.5   & 0.674 \\ 
\#161   & -0.3   & -2.8   & 4000   & -3.5   & 8000   & 9.5   & 0.640 \\ 
\#162   & -0.3   & -2.7   & 4000   & -3.5   & 8000   & 9.5   & 0.606 \\ 
\#163   & -0.3   & -1.5   & 7000   & -3.5   & 7800   & 8.5   & -0.758 \\ 
\#164   & -0.3   & -1.5   & 7000   & -3.5   & 8000   & 8.5   & -0.816 \\ 
\#165   & -0.3   & -1.5   & 8000   & -3.5   & 8500   & 9.5   & 0.234 \\ 
\#166   & -0.3   & -1.5   & 8000   & -4.0   & 8500   & 9.5   & 0.227 \\ 
\end{longtable}
}

\end{appendix}

\end{document}